\documentclass[iop]{emulateapj}

\def\msun{\hbox{M$_{\odot}$}}
\def\mdot{\hbox{$\dot M$}}
\def\micron{$\mu$m}
\def\microns{$\mu$m}
\def\lsun{\rm L_{\sun}}
\def\rsun{$\mbox{R}_{\odot}$}

\newcommand\be{\begin{equation}}
\newcommand\en{\end{equation}}

\begin{document}

\shortauthors{Espaillat et al.}
\shorttitle{The Transitional Disk Class}

\title{On the Transitional Disk Class: Linking Observations 
of T Tauri Stars \& Physical Disk Models}

\author{C. Espaillat\altaffilmark{1,2},
L. Ingleby\altaffilmark{3}, 
J. Hern\'{a}ndez\altaffilmark{4}, 
E. Furlan\altaffilmark{5,6}, 
P. D'Alessio\altaffilmark{7}, 
N. Calvet\altaffilmark{3},
S. Andrews\altaffilmark{1}, 
J. Muzerolle\altaffilmark{8}, 
C. Qi\altaffilmark{1}, \&
D. Wilner\altaffilmark{1}
}

\altaffiltext{1}{NSF Astronomy \& Astrophysics Postdoctoral Fellow}
\altaffiltext{2}{Harvard-Smithsonian Center for Astrophysics, 60 Garden
Street, MS-78, Cambridge, MA, 02138, USA; cespaillat@cfa.harvard.edu,
sandrews@cfa.harvard.edu, cqi@cfa.harvard.edu, dwilner@cfa.harvard.edu}
\altaffiltext{3}{Department of
Astronomy, University of Michigan, 830 Dennison Building, 500 Church
Street, Ann Arbor, MI 48109, USA; lingleby@umich.edu, ncalvet@umich.edu}
\altaffiltext{4}{Centro de Investigaciones de Astronom\'{i}a (CIDA),
Merida, 5101-A, Venezuela; jesush@cida.ve} 
\altaffiltext{5}{Visitor at the Infrared
 Processing and Analysis Center, California Institute
 of Technology, 770 S. Wilson Ave., Pasadena, CA 91125, USA}
\altaffiltext{6}{National Optical Astronomy Observatory, 950 N. Cherry
 Ave., Tucson, AZ 85719, USA; Elise.Furlan@jpl.nasa.gov} 
\altaffiltext{7}{Centro de Radioastronom\'{i}a y Astrof\'{i}sica,
Universidad Nacional Aut\'{o}noma de M\'{e}xico, 58089 Morelia,
Michoac\'{a}n, M\'{e}xico; p.dalessio@crya.unam.mx}
\altaffiltext{8}{Space Telescope Institute, 3700 San Martin Drive,
Baltimore, MD 21218, USA; muzerol@stsci.edu}

\begin{abstract} 

Two decades ago ``transitional disks''  described spectral energy
distributions (SEDs) of T Tauri stars with small near-IR excesses, but significant mid- and far-IR
excesses.  Many inferred this indicated dust-free holes in disks,
possibly cleared by planets. Recently, this term has been applied
disparately to objects whose {\it Spitzer} SEDs diverge from the
expectations for a typical full disk.  Here we use irradiated accretion
disk models to fit the SEDs of 15 such disks in NGC~2068 and IC~348. One
group has a ``dip'' in infrared emission while the others'  continuum
emission decreases steadily at all wavelengths.  We find that the former
have an inner disk hole or gap at intermediate radii in the disk and we
call these objects ``transitional'' and ``pre-transitional'' disks,
respectively.  For the latter group, we can fit these SEDs with full
disk models and find that millimeter data are necessary to break the
degeneracy between dust settling and disk mass. We suggest the term
``transitional'' only be applied to objects that display evidence for a
radical change in the disk's radial structure. Using this definition, we
find that transitional and pre-transitional disks tend to have lower mass accretion
rates than full disks and that transitional disks have lower accretion
rates than pre-transitional disks. These reduced accretion rates onto
the star could be linked to forming planets. Future observations of
transitional and pre-transitional disks will allow us to better quantify the signatures
of planet formation in young disks.

\end{abstract}


\section{Introduction} \label{intro}

Disks around T Tauri stars (TTS) are thought to be the sites of planet
formation. However, many questions exist concerning how the gas and dust
in the disk evolve into a planetary system and observations of TTS may
provide clues. There are some objects in particular that have gained
increasing attention in this regard. 
Over two decades ago, \citet{strom89} detected ``possible evidence of changes in disk structure with time''
as evidenced by ``small near-IR excesses, but significant mid- and far-IR
excesses'' indicating ``inner holes'' in disks.  Those authors proposed that these objects were
``{\it in transition} from massive, optically thick structures that extend
inward to the stellar surface, to low-mass, tenuous, perhaps post-planet-building
structures.''

Usage of the term ``transitional disk'' gained substantial momentum in the literature
after it was used by \citet{calvet05} to describe disks with inner holes using data
from the {\it Spitzer Space Telescope}'s \citep{werner04} 
Infrared Spectrograph \citep[IRS;][]{houck04}.
Before {\it Spitzer},
the spectral energy distributions
(SEDs) used to identify disks with inner holes were
based solely on near-infrared (NIR)
ground-based photometry and IRAS mid-IR (MIR) photometry \citep{strom89,
skrutskie90}.
{\it Spitzer} allowed
the opportunity to study these objects in greater detail.
The unprecedented resolution and simultaneous wavelength 
coverage ($\sim$5 and 38~{\microns}) of {\it Spitzer} IRS uncovered new
details regarding these disks \citep{dalessio05, calvet05, furlan06}.  Some SEDs had nearly
photospheric NIR (1--5~{\micron}) and MIR (5--20~{\micron}) emission,
coupled with substantial emission above the stellar photosphere at
wavelengths beyond $\sim$20~{\micron}.  Others had significant NIR
excesses relative to their stellar photospheres, but still exhibited MIR
dips and substantial excesses beyond $\sim$20~{\micron}.

Detailed modeling of many of the above-mentioned SEDs has been
performed. SEDs of disks with little or no NIR and MIR emission have
been fit with models of inwardly truncated optically thick disks
\citep{calvet02,calvet05,espaillat07a,espaillat08b}.  The inner edge or
``wall'' of the outer disk is frontally illuminated by the star,
dominating most of the emission seen in the IRS spectrum.   In this
paper, we refer to these objects with holes in their dust distribution
as transitional disks (TD).  Some of the holes in TD are relatively
dust-free \citep[e.g. DM Tau;][]{calvet05, espaillat10}  while  SED
model fitting indicates that others with strong 10{\micron} silicate
emission have a small amount of optically thin dust in their disk holes
to explain this feature \citep[e.g. GM Aur;][]{calvet05, espaillat10}. 
For SEDs with substantial NIR emission accompanied by a MIR dip, we can
fit the observed SED with an optically thick inner disk separated by an
optically thin gap from an optically thick outer disk
\citep{espaillat07b}.  Here we call these pre-transitional disks (PTD). 
Like the TD, we see evidence for relatively dust-free gaps \citep[e.g.
UX Tau A;][]{espaillat07b, espaillat10} as well as gaps with some small,
optically thin dust to explain strong 10{\micron} silicate emission
features \citep[e.g. LkCa 15;][]{espaillat07b, espaillat10}. For many
TDs and PTDs, the truncation of the outer disk has been confirmed with
sub-millimeter and millimeter interferometric imaging \citep[e.g.
DM~Tau, GM~Aur, UX~Tau~A,
LkCa~15;][]{hughes07,hughes09,andrews09,andrews10,andrews11} as well as
NIR imaging \citep[i.e. LkCa~15;][]{thalmann10}.  
In a few cases, the optically thick inner disk of PTD has been confirmed using the ``veiling\footnote{
``Veiling'' occurs when an excess continuum \citep{hartigan89} ``fills
in'' absorption lines, causing them to appear significantly weaker than
the spectrum of a standard star of the same spectral type
\citep{hartigan89}. The veiling observed in pre-transitional disks is
similar to that observed in full disks where the veiling has been
explained by emission from the inner disk edge or ``wall'' of an
optically thick disk \citep{muzerolle03}. }" of near-infrared spectra
\citep{espaillat08a, espaillat10}
and near-infrared interferometry has confirmed 
that the inner disk is small \citep{pott10,olofsson11}.

The distinct SEDs of TD and PTD most likely signify that these objects
are being caught in an important phase in disk evolution.  Many
researchers have posited that these disks are forming planets on the
basis that cleared dust regions are predicted by planet formation models
\citep[e.g.][]{paardekooper04, zhu11,dodson11}.  Recently, a potential
protoplanet has been reported in the pre-transitional disk around LkCa~15
\citep{kraus11b} as well as around T Cha \citep{huelamo11}.
Stellar companions can
also clear the inner disk \citep{artymowicz94} but many stars harboring
transitional disks are single stars \citep{kraus11}.  Even in cases
where stellar-mass companions have not been ruled out, the large holes
and gaps observed are most likely evidence of dynamical clearing. 
Photoevaporation cannot explain disks with large cavities and high mass
accretion rates \citep{owen11} and dust evolution alone can not explain
the sharp decreases in surface density seen in the SED and
interferometric visibilities.

Given the potential link between disks with gaps and holes and planet
formation, interest in transitional disks has grown. Some studies have
focused on further understanding the behavior of the currently known
members in this class of objects and have discovered IR variability
\citep[][E11]{muzerolle10, espaillat11}. Other studies have taken a
broader approach, working towards expanding the known number of disks
undergoing clearing \citep[e.g.][]{lada06, hernandez07, cieza10,
muzerolle10, luhman10, merin10, currie11}. As the literature on {\it
Spitzer} observations of TTS expands, so does the terminology applied to
disks around TTS. These disks are referred to as primordial, full,
transitional, pre-transitional, kink, cold, anemic, homologously
depleted, classical transitional, weak excess, warm excess, and evolved.
  However, these terms are not applied consistently. The issue is
highlighted with the discrepancies in the reported fractions of
transitional disks and disk clearing timescales \citep{merin10,
luhman10, muzerolle10, currie11, hernandez10}.  In many cases, the data
is the same; the differences arise from nomenclature.

If the goal is to better understand disk evolution, it is important to
look past the nomenclature and decipher the underlying disk structure we
can infer from the observations, while paying special attention to the
limitations of both the observations and the tools we use to interpret
them, namely disk models. 
To this end, we chose a sample for our study which incorporated 
objects whose SEDs are similar to those that have been referred to as ``transitional''
in the literature.
We focus on 15 disks in NGC~2068 and IC~348
with {\it Spitzer} IRS spectra.  NGC~2068 and IC~348 are older
\citep[$\sim$2~Myr and $\sim$3~Myr, respectively;][]{flaherty08,luhman03}
and more clustered star-forming regions
with higher extinction than Taurus, where most detailed studies of
transitional disks have focused.  According to our definitions stated
above, our sample contains five TD, five PTD, and five objects with
decreasing emission at all IR wavelengths (i.e. negative MIR slopes)
which we classify as full disks.  Out of the objects reported by
\citet{muzerolle10}, this includes 5 of the 15 transitional and pre-transitional disks
in IC~348 and 4 of the 6 transitional and pre-transitional disks in NGC~2068.  One of
the objects classified as a full disk in that work is classified as a
PTD here. 

In Section 2, we present optical photometry and spectroscopy,
infrared spectra, and millimeter flux densities that we used to compile
SEDs for our objects, derive stellar properties, and measure mass
accretion rates. In Section 3, we fit the SEDs of our objects with
irradiated accretion disk models and in Section 4 we discuss the
limitations of the models and the observations, disk semantics, and the
mass accretion rates of transitional and pre-transitional disks.

\section{Observations \& Data Reduction} \label{redux}

Before conducting modeling, we first compiled SEDs and mass accretion
rates for the targets in our sample (Table~\ref{tab:targetlog}).   To
supplement the existing information in the literature, we collected
optical photometry for all of our IC~348 objects, optical spectra for
three of the IC~348 targets and all of our NGC~2068 targets, and
millimeter flux densities for two of our IC~348 targets.  We also
present infrared spectra for all the targets in this paper.  Below we
discuss the details of the observations and data reduction.

\subsection{Optical Photometry}

During November 19--23, 2011 we used the 4K CCD imager on the 1.3-m
McGraw-Hill
telescope\footnote{http:$/$$/$www.astronomy.ohio-state.edu$/$$\sim$
jdeast$/$4k$/$} of the MDM Observatory to obtain UVRI photometry of the
IC348 stellar cluster. The FOV of the instrument is
21$^\prime$$\times$21$^\prime$ (0.315$^\prime$$^\prime$ per unbinned
pixel).  For these observations, we used 2$\times$2 binning
(0.62$^\prime$$^\prime$ per pixel).  The flat-field, overscan, and
astrometric calibration were performed using an IDL program written by
Jason
Eastman\footnote{http://www.astronomy.ohio-state.edu$/$$\sim$jdeast$/$4k
$/$proc4k.pro} specifically designed for the 4K CCD imager. Since large
electronic structures are not stable enough to be reliably subtracted,
we did not apply corrections using two dimensional biases. The
photometric calibration of all images was carried out using the standard
procedure and the {\it daophot} and {\it photcal} packages in IRAF, with
standard stars selected from \citet{landolt92}.  Photometry is presented
in Table~\ref{tab:phot}.

\subsection{Optical Spectroscopy}

\begin{figure}
\epsscale{1}
\plotone{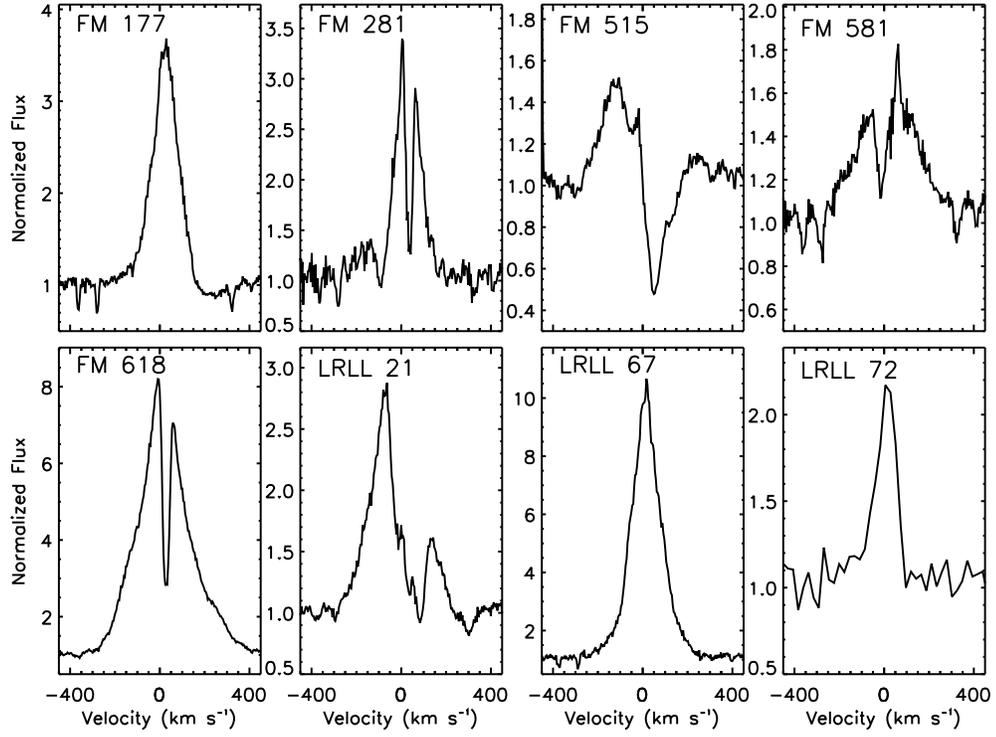}
\caption[]{
High-resolution H$_{\alpha}$ line profiles obtained using the MIKE
spectrograph for three of our IC~348 targets and all five of our
NGC~2068 targets.  Wide and asymmetric profiles (i.e.  FM~515, FM~581,
FM~618,  LRLL~21) indicate substantial accretion onto the star and gas
in the inner disk while narrow profiles indicate low accretion rates
onto the star \citep[e.g.][]{white03}.  We note that the spectrum of
LRLL~72 has been binned up by a factor of $\sim$5 here for clarity.
}
\label{figmdot}
\end{figure}

In order to obtain mass accretion rate estimates (Section~\ref{sec:mdot}
and Figure~\ref{figmdot}), we collected MIKE double-echelle spectrograph
\citep{bernstein03} data on the 6.5m Magellan Clay telescope for all of
our targets in NGC~2068 (FM~177, FM~281, FM~515, FM~581, and FM~618) and
some of our objects in IC~348 (LRLL~21, LRLL~67, and LRLL~72).  The
observations for the NGC~2068 and IC~348 objects were taken on February
10-11, 2007 and January 19, 2009, respectively. We used a slit size of
0.7$^\prime$$^\prime$$\times$5.0$^\prime$$^\prime$ and 2$\times$2 pixel
on-chip binning with exposure times of 800-1200s. Data were reduced
using the MIKE data reduction
pipeline.\footnote{http:$/$$/$web.mit.edu$/$$\sim$burles$/$www$/$MIKE$/$
}

\subsection{Infrared Spectroscopy}

Here we present {\it Spitzer} IRS spectra for each of our targets.
Spectra for NGC~2068 and IC~348 were obtained in Program 58 (PI: Rieke)
and Program 2 (PI: Houck), respectively. All of the observations were
performed in staring mode using the IRS low-resolution modules,
Short-Low (SL) and Long-Low (LL), which span wavelengths from
5--14~{\microns} and 14-38~{\microns}, respectively, with a resolution
$\lambda/\delta\lambda\sim$90.

Details on the observational techniques and general data reduction steps
can be found in \citet{furlan06} and  \citet{watson09}.  We provide a
brief summary. Each object was observed twice along the slit, at a third
of the slit length from the top and bottom edges of the slit. Basic
calibrated data (BCD) with pipeline version S18.7 were obtained from the
{\it Spitzer} Science Center.  With the BCDs, we extracted and
calibrated the spectra using the SMART package \citep{higdon04}. Bad and
rogue pixels were corrected by interpolating from neighboring pixels.

Most of the data were sky subtracted using optimal extraction
\citep{lebouteiller10}. The exceptions were LRLL~37, LRLL~55, and
LRLL~68. In LRLL~37, there was an artificial structure in the
5--8~{\micron} region which was removed by performing off-nod sky
subtraction. A similar structure was seen in LRLL~55 and LRLL~68, but
due to the high background in the area, optimal extraction was necessary
for the LL order.  Therefore, the final spectra for LRLL~55 and LRLL~68
are a combination of the off-nod sky-subtracted SL spectra and the
optimally extracted LL spectra.

\begin{figure}
\epsscale{1}
\plotone{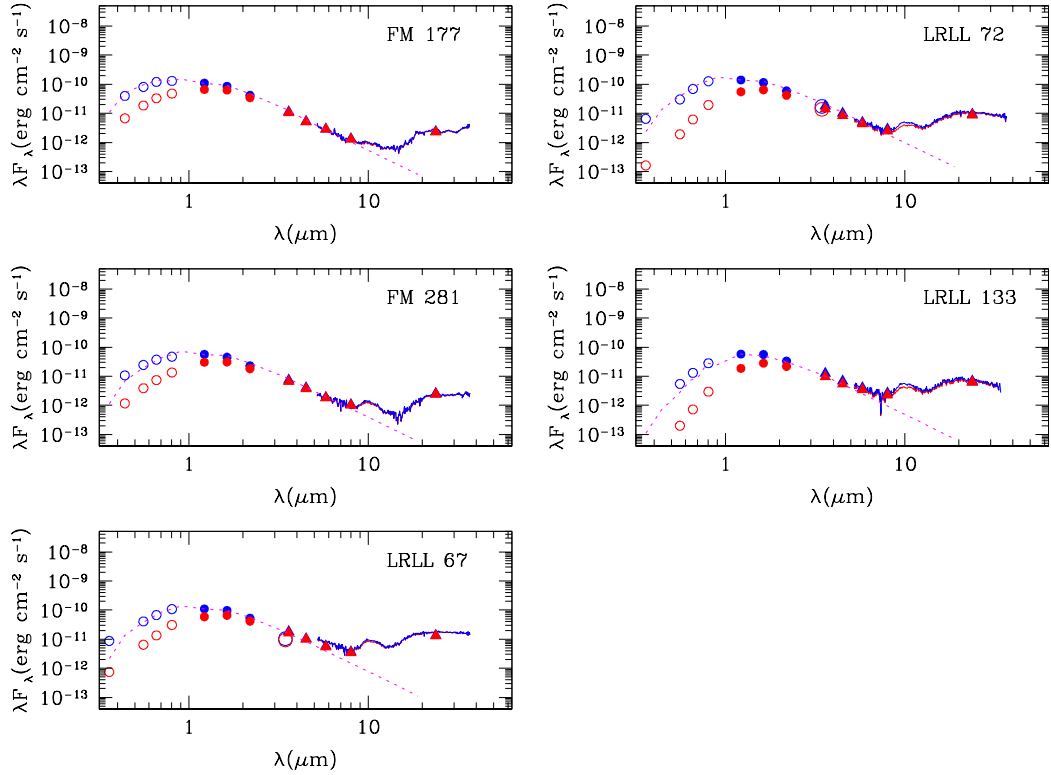}
\caption[]{
SEDs of transitional disks in our sample.  We show both the observed
fluxes (red) and dereddened fluxes (blue; see Table~\ref{tab:prop} for
A$_V$). In the NIR the emission is similar to that of the stellar
photosphere (broken magenta line; \citet{kh95}), but rises in the MIR
and at longer wavelengths. This indicates that the dust in the inner
disk has been removed and that there is a hole in the disk. Open circles
correspond to ground-based U-, B-, V-, R-, I-, and L-band photometry;
closed circles are 2MASS J-, H-, and K-band photometry; triangles are
{\it Spitzer} IRAC and MIPS photometry. See Section~\ref{sec:starprop}
for data references. Solid lines are {\it Spitzer} IRS spectra from this
work. (A color version of this figure is available in the online
journal.)
}
\label{figsedtd}
\end{figure}

\begin{figure}
\epsscale{1}
\plotone{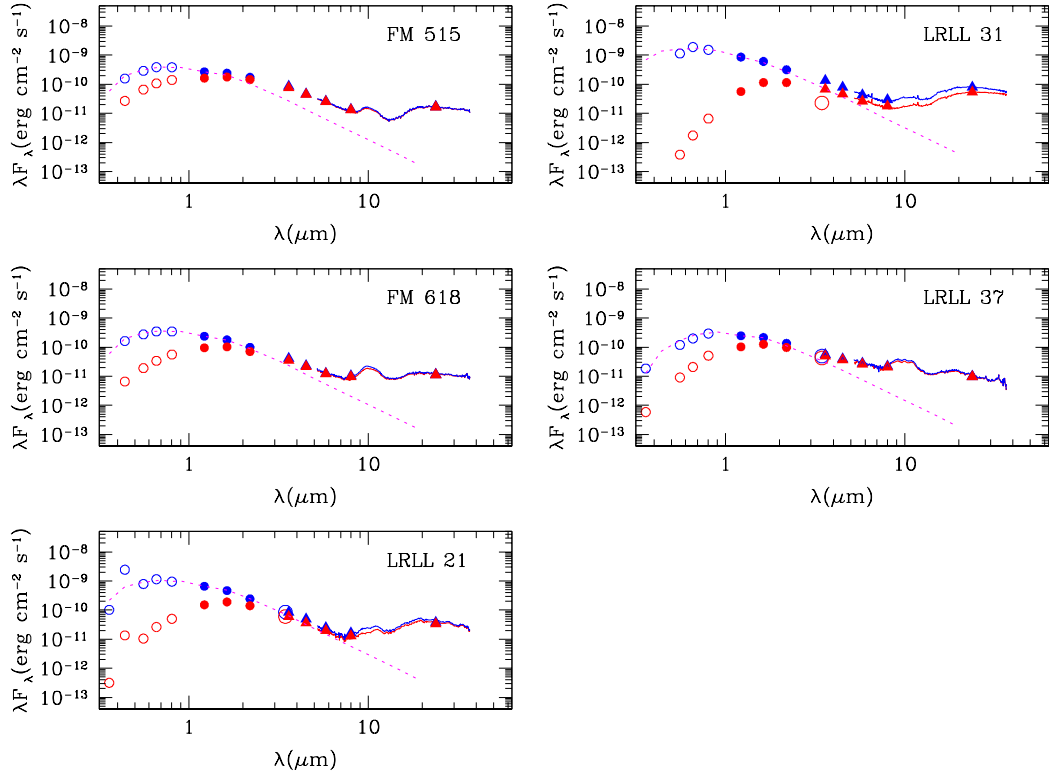}
\caption[]{
SEDs of pre-transitional disks in our sample.  For LRLL~31, FM~515, and
FM~618, there is significant NIR emission above the stellar photosphere,
but there is also a dip in the MIR emission and the flux increases at
longer wavelengths.  This indicates a gap at intermediate radii in the
disk.  In the case of LRLL~37, there is no clear ``dip'' in the
emission;  this object was classified as a pre-transitional disk based
on its strong silicate emission which could not be reproduced with a
full disk model (see Section~\ref{sec:modsed}). 
We classify LRLL~21 as a PTD here even though its NIR excess is weak in 
this epoch;
\citet{flaherty12}
find that LRLL~21 has significant NIR blackbody emission in more recent IRS
observations.
The color scheme,
symbols, and lines are the same as in Figure~\ref{figsedtd}. (A color
version of this figure is available in the online journal.)
}
\label{figsedptd}
\end{figure}

\begin{figure}
\epsscale{1}
\plotone{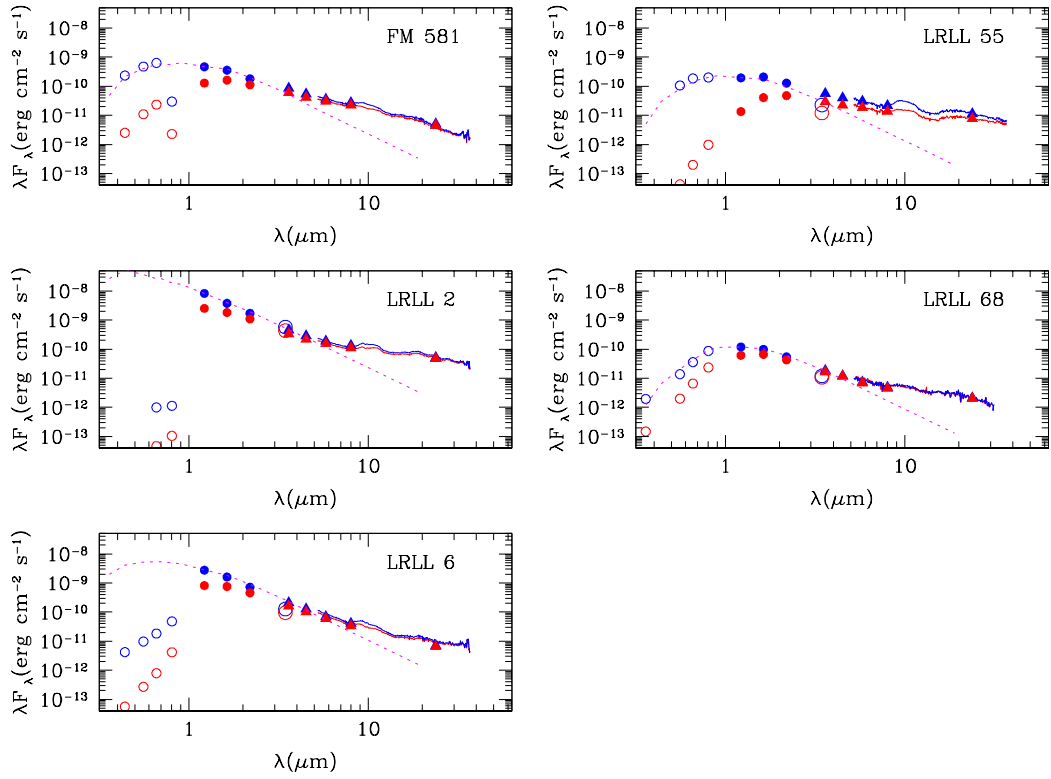}
\caption[]{
SEDs of full disks in our sample.  The emission decreases at all
wavelengths. The color scheme, symbols, and lines are the same as those
used in Figure~\ref{figsedtd}. (A color version of this figure is
available in the online journal.)
}
\label{figseded}
\end{figure}

To flux calibrate the observations we used a spectrum of $\alpha$ Lac
(A1 V).  We performed a nod-by-nod division of the target spectra and
the $\alpha$ Lac spectrum and then multiplied the result by a template
spectrum \citep{cohen03}. The final spectrum was produced by averaging
the calibrated spectra from the two nods. Our spectrophotometric
accuracy is 2--5$\%$ estimated from half the difference between the
nodded observations, which is confirmed by comparison with IRAC and MIPS
photometry. We note that there are artifacts in the spectra of LRLL~68
and LRLL~133 beyond $\sim$30~{\micron}. For clarity, we manually trim
the spectra to exclude these regions. The final spectra used in this
study are shown in
Figures~\ref{figsedtd},~\ref{figsedptd},~and~\ref{figseded}.

\subsection{Millimeter Flux Densities}

We observed LRLL~21, LRLL~31, LRLL~67, LRLL~68, and LRLL~72 in IC~348
with the Submillimeter Array (SMA) on November 12, 2008. We used the
Compact Configuration with six of the 6 meter diameter antennas at 345
GHz (860~{\micron}) with a full correlator bandwidth of 2 GHz.
Calibration of the visibility phases and amplitudes was achieved with
observations of the quasars 3C~111 and 3C~84, typically at intervals of
20 minutes. Observations of Uranus provided the absolute scale for the
flux density calibration. The data were calibrated using the MIR
software
package.\footnote{http://www.cfa.harvard.edu/$\sim$cqi/mircook.html}  We
detected LRLL~31 and LRLL~67 with flux densities of 0.062$\pm$0.006~Jy
and 0.025$\pm$0.011~Jy, respectively.  We did not detect LRLL~21,
LRLL~68, or LRLL~72 and measure a 3${\sigma}$ upper limit of 0.015~Jy
for these objects.

\section{Analysis}

Here we model the SEDs of the targets in our sample.  First we collect
the stellar properties of our objects, either by adopting literature
values or deriving our own in Section~\ref{sec:starprop}. These stellar
properties are important input parameters for our physically motivated
models which we discuss in Section~\ref{sec:diskmodel}.  In
Section~\ref{sec:modsed} we discuss the results of our SED model
fitting, as well as the degeneracies that exist given that we lack
millimeter observations for many of our targets.

\subsection{Stellar Properties} \label{sec:starprop}

Stellar parameters are listed in Table~\ref{tab:prop}. M$_{*}$ was
derived from the HR diagram and the \citet{siess00} evolutionary tracks
using T$_{*}$ and L$_{*}$. Stellar temperatures are from \citet{kh95},
based upon the spectral types adopted for the targets in NGC ~2068 and
IC~348 \citep[][respectively]{flaherty08,luhman03}. Luminosities are
calculated with dereddened J-band photometry following \citet{kh95}
assuming a distance of 400~pc for NGC~2068 \citep{flaherty08} and 315~pc
for IC~348 \citep{luhman03}. R$_{*}$ is calculated using the derived
luminosity and adopted temperature. The derivation of the mass accretion
rates and accretion luminosities are discussed in
Section~\ref{sec:mdot}.

All photometry for our NGC~2068 objects is taken from
\citet{flaherty08}.  This includes BVRI, 2MASS JHK, IRAC, and MIPS data.
 We note that  \citet{flaherty08} present photometry from the SDSS
survey, which we convert to Johnson-Cousins BVRI following
\citet{jordi06}. IRAC and MIPS photometry for IC~348 comes from
\citet{lada06} and 2MASS JHK data is from \citet{skrutskie06}.  UBVRI
photometry for our IC~348 targets comes mainly from this work, but is
supplemented by values in the literature. All UBVRI photometry for
LRLL~31, LRLL~55, LRLL~67, LRLL~68, LRLL~72, and LRLL~133 are solely
from this work.  UVR data for LRLL~21 and LRLL~37 are from this work,
but we use I-band magnitudes from \citet{luhman03} for both and a B-band
magnitude for LRLL~21 from \citet{herbig98}.  We use  R and I data from
\citet{herbig98} for LRLL~2.  BVRI data for LRLL~6 is also from
\citet{herbig98}.  All L-band magnitudes are from \citet{haisch01a}.

Extinctions were measured by comparing V-R, V-I, R-I, and I-J colors to
photospheric colors from \citet{kh95}.  We used the \citet{mathis90}
extinction law for objects with A$_V$$<$ 3.  For A$_V$$\geq$ 3, we use
the \citet{mcclure09} extinction law.  We adopt R$_V$=5 which is more
appropriate for the denser regions studied here \citep{mathis90}.  In
most cases, extinctions based on I-J colors gave the best fit.  Since we
have no I-band data for FM~581 we adopt an extinction based on V-R; 
LRLL~2 and LRLL~6 have no VRI photometry in the literature and so we
adopt the extinction measured by \citet{luhman03}.  All extinctions used
in this work are listed in Table~\ref{tab:prop}. We note that most of
our extinctions are similar to those in the literature
\citep{luhman03,flaherty08}.  In cases where we found differences, we
chose to rely on our measurements since they are based on I-J colors
which have recently been shown to be the least affected by excess
emission at shorter wavelengths \citep[][McClure et al., in
preparation]{fischer11}.  For early-type stars, the peak of the stellar
emission would be at shorter wavelengths.  However, the early-type star
in our sample (LRLL~2) has bad photometry so we do not explore this
point further.

\subsubsection{Accretion Rates} \label{sec:mdot}

During the classical TTS phase of stellar evolution, young objects
accrete material from the disk onto the star via magnetospheric
accretion \citep{uchida84}.  The infalling gas impacts the stellar
surface at approximately the free fall velocity creating a shock which
heats the gas to $\sim$ 1~MK \citep{calvet98}.  The shock emission is
reprocessed in the accretion column and the observed spectrum peaks in
the ultraviolet \citep{calvet98}.  The best estimate of the accretion
rate is found by measuring the total luminosity emitted in the accretion
shock, i.e. the accretion luminosity.  While ultraviolet emission is
difficult to observe from ground-based observatories,  there are many
tracers of the accretion luminosity at longer wavelengths (see
\citet{rigliaco11} for a comprehensive list).  A few of those tracers
include excess emission observed in U-band photometry, emission in the
NIR Ca II triplet lines, and the H$\alpha$ line profile.

Second to the ultraviolet excess, U-band excesses are the best measure
of the flux produced in the shock and have been shown to correlate with
the total shock excess \citep{calvet98}.  However, at low $\mdot$
emission at $U$ may be dominated by chromospheric emission
\citep{houdebine96,franchini98};  this chromospheric excess can confuse
determinations of the accretion rate \citep{ingleby11}.  While possible
from the ground, U-band observations are still difficult to obtain,
especially when extinction towards the source is relatively high as in the case of
our sample, all with $A_{V}>1$.  Observations of the Ca II near-infrared
triplet are easily obtained from the ground, even for high $A_V$
sources, and the flux in the 8542 {\AA} line also correlates with the
accretion luminosity \citep{muzerolle98}.  Ca II is observed in emission
in accreting sources but is unreliable at low accretion rates, when the
chromospheric emission rivals that from accretion in strength
\citep{yang07,batalha93, ingleby11b}. H$\alpha$ is commonly used as a
tracer of accretion, both by measuring the equivalent width and the
velocity width of the line in the wings
\citep{white03,barrado03,natta04}.  Models of magnetospheric accretion
have reproduced the observed velocities in H$\alpha$, tracing material
traveling at several hundred km~s$^{-1}$ near the accretion shock
\citep{lima10,muzerolle03}.

The mass accretion rates adopted for our sample are listed in
Table~\ref{tab:prop}. We used U-band photometry from this work and the
relation in \citet{gullbring98} to measure mass accretion rates for
LRLL~37 and LRLL~68 and an upper limit for LRLL~133.\footnote{We measured
an upper limit for LRLL~55 of 4$\times$10$^{-6}$ ${\msun}$
yr$^{-1}$  using U-band photometry.   However, this very high upper limit does not provide
useful constraints for the purposes of this paper and so we do not comment on it further.} For FM~177, FM~281,
FM~515, FM~581, FM~618, LRLL~21, LRLL~67, and LRLL~72,  we used high
resolution echelle spectra obtained with MIKE to measure the width of
the H$\alpha$ line in the wings (at 10\% of the maximum flux;
Figure~\ref{figmdot}).  We then compared this to the relation between
line width and $\mdot$ in \citet{natta04} to obtain mass accretion rate
estimates for these eight sources.  While our MIKE spectra covered both
H$\alpha$ and the Ca II triplet, H$\alpha$ provided a more accurate
estimate of $\mdot$ due to confusion with chromospheric Ca II emission
at the levels of accretion found in these sources.  Given the
chromospheric appearance of the H$\alpha$ profile of LRLL~72, its mass
accretion rate should be taken as an upper limit.  For LRLL~31 we
adopted a mass accretion rate from the literature. We do not have mass
accretion rate measurements for LRLL~ 2 or LRLL~6.

For NGC 2068, we compared our derived accretion rates to those in
\citet{flaherty08} who calculated the amount of excess continuum
emission necessary to produce the observed veiling of the photospheric
lines. Within a factor of 2--3, the normal error in $\mdot$ estimations,
both calculations of the accretion rate agree, with a few exceptions. 
When comparing the H$\alpha$ line widths at 10\% we find that our MIKE
line profile of FM~281 is $\sim180\;\rm{km\;s^{-1}}$ narrower than when
observed by \citet{flaherty08}.  In addition, our observation of FM~177
is consistent with an accreting source, while when observed by
\citet{flaherty08} its H$\alpha$ profile was consistent with that of
chromospheric emission.  Variability is known to occur in T Tauri
stars so the decrease in H$\alpha$ and $\mdot$ are not unexpected
\citep{cody10}.  For these objects we chose to adopt the accretion rate
obtained from our MIKE observations.  The biggest uncertainty in
calculating accretion rates using veiling is the choice of bolometric
correction, which can vary in value by a factor of 10 depending on which
analysis is used \citep{white04} and the spectrum of the excess emission
which veils the photospheric lines can be complicated \citep{fischer11}.

\subsection{Disk Model} \label{sec:diskmodel}

We try to reproduce the SEDs presented in
Figures~\ref{figsedtd},~\ref{figsedptd},~\&~\ref{figseded} using the
irradiated accretion disk models of
\citet{dalessio98,dalessio99,dalessio01,dalessio05,dalessio06}. We point
the reader to those papers for details of the model and to
\citet{espaillat10} for a summary of how we apply the model to the SEDs
of transitional and pre-transitional disks.  Here we provide a brief
review of the salient points of the above works.

When we refer to a ``full disk model'' we mean a disk model composed of
an irradiated accretion disk and a frontally illuminated wall at the
inner edge of the disk which is located at the dust sublimation radius. 
The inner wall dominates the emission in the NIR, the wall and disk both
contribute to the MIR emission, and the disk dominates the emission at
longer wavelengths. Compared to a full disk model, a pre-transitional
disk model has a gap within the disk. In this case, we include a
frontally illuminated wall at the dust sublimation radius and another
wall at the gap's outer edge. For this outer wall we include the shadow
cast by the inner wall \citep{espaillat10}. We do not include an inner
irradiated accretion disk behind the inner wall since previous work has
shown that the inner wall dominates the emission at these shorter
wavelengths \citep{espaillat10}. Behind the outer wall we include an
irradiated accretion disk in cases where we have millimeter data, which
is necessary to constrain the outer disk. The inner wall dominates the
NIR emission while the outer wall dominates the emission from
$\sim$20--30~{\micron}. The outer disk dominates the emission beyond
$\sim$40~{\micron}. A transitional disk model is very similar to that of
a pre-transitional disk model except that we do not include an inner
wall at the dust sublimation radius. In some instances, we include a
small amount of optically thin dust in the inner hole or gap in
transitional and pre-transitional disks to reproduce the 10~{\micron}
silicate emission feature. We calculate the emission from this optically
thin dust region following \citet{calvet02}.

\subsubsection{Disk Properties}

Table~\ref{tab:disk} lists the model-derived properties of our sample. 
The heights of the inner and outer walls (z$_{wall}$) and the maximum
grain sizes (a$_{max}$) are adjusted to fit the SED.  $T_{wall}$ is the
temperature at the surface of the optically thin wall atmosphere. The
temperature of the inner wall of full disks and pre-transitional disks
(T$_{wall}^i$) is held fixed at 1400~K (except for FM~515, see
Section~\ref{sec:modsed}) which is the typical temperature of dust at
the sublimation radius \citep{muzerolle03}. The temperature of the outer
wall (T$_{wall}^o$) in transitional and pre-transitional disks is varied
to fit the SED, particularly the IRS spectrum.  The radius of the wall
($R_{wall}$) is derived using the best fitting $T_{wall}$ following
Equation 2 in \citet{espaillat10}.

Previously, we have seen that in transitional and pre-transitional
disks, the  IRS spectrum is dominated by the outer wall while the outer
disk dominates the millimeter emission (E11). Since the majority of the
emission seen by IRS is from the outer wall, the IRS spectrum is a good
constraint of the hole$/$gap size and in most cases SED-derived
hole$/$gap sizes are in reasonable agreement with those obtained with
millimeter imaging \citep{andrews11}. For objects in this work with no
millimeter data, we do not include an outer disk given that its
contribution to the IRS SED is expected to be small. We did include an
outer disk for the two objects in the sample for which we have
millimeter data: the transitional disk LRLL~67 and the pre-transitional
disk LRLL~31. We also included an outer disk when modeling each of our
full disk targets since additional emission from an outer disk is
necessary to reproduce the observed MIR emission. The parameters of the
outer disk which are varied to fit the SED are the viscosity parameter
($\alpha$) and the settling parameter ($\epsilon$; see
Section~\ref{dustprop}).  One can interpret varying $\alpha$ as fitting
for the disk mass since M$_{disk}\propto{\mdot}/{\alpha}$ \citep[see
Equation 38 in][]{dalessio98}.  
We note that the mass accretion rate onto the star does not necessarily reflect the mass 
transport across the outer disk, especially in the case of TD and PTD where the mass
accretion rate onto the star is likely an underestimate of the mass transport
across the outer disk (see Section~\ref{tdmdot}).
We also do not expect that the mass accretion rate is constant throughout the disk.
However, for simplicity, here we assume that the mass accretion rate measured onto the 
star is representative of the disk's accretion rate.

We will discuss how the lack of millimeter constraints leads to
degeneracies in our outer disk model fitting in
Section~\ref{sec:modsed}. There we also discuss the effect the adopted
disk inclination and outer radius have on the simulated SED. We assume
that the inclination of the disk is 60$^{\circ}$ for all of our objects
and that they have an outer disk radius of 300~AU (except FM~581, see
Section~\ref{sec:modsed}).

\subsubsection{Dust Properties} \label{dustprop}

The opacity of the disk, and hence the temperature structure and
resulting emission, is controlled by dust.  The dust opacity depends on
the composition of the dust assumed.  It also depends on changes in the
dust due to grain growth and settling. Grains grow through collisional
coagulation and settle to the disk midplane due to gravity. Since in
this work we include models of several full disks, here we review the
effect that the dust properties have on the disk structure in more
detail following \citet{dalessio06}.

{\it Dust Settling.} The settling of dust has important, and often
overlooked, effects on the disk's density--temperature distribution and
emission. When there is some degree of settling, the dust-to-gas mass
ratio of grains in the disk atmosphere decreases with respect to the
standard value (i.e. the diffuse interstellar medium). This has several
effects: (1) it decreases the opacity of the upper layers; this allows
the impinging external radiation to penetrate deeper into the disk,
decreasing the height of the irradiation surface\footnote{The height of
the disk irradiation surface, $z_s$, is defined by the region where
$\tau_s$, the radial optical depth to the stellar radiation, is $\sim
1$.} and making it geometrically flatter \citep[see Figure 3
in][]{dalessio06}, which in turn decreases the fraction of the
irradiation flux intercepted by this surface,\footnote{As stellar
radiation enters the disk, it does so at an angle to the normal of the
disk surface ($\theta_0=cos^{-1}\mu_0$).  A fraction of the stellar
radiation is scattered and the stellar radiation captured by the disk is
$\sim( \sigma T_*^4/\pi) (R_*/R)^2 \mu_0$ (see \citet{calvet91} for
further discussion). Therefore, if the disk is more flared, $\mu_0$ is
larger and more stellar irradiation will be intercepted by the disk and
it will be hotter and emit more radiation.} decreasing the continuum
flux emerging from the disk, (2) since most of the external radiation is
deposited at the irradiation surface, lowering it changes the
temperature-density structure of the atmospheric layers, where the
temperature inversion occurs (also called the super-heated layers),
modifying their contribution to the SED, and finally, (3) it changes the
emissivity of the disk interior; in this region the dust-to-gas mass
ratio of the grains increases given that the grains removed by depletion
from the upper layers are now located deeper in the disk.

Some of the above-mentioned effects can be accounted for by arbitrarily
changing the disk surface height as a function of radius
\citep[e.g.][]{miyake95, currie11, sicilia11}, and this will probably
give a reasonable estimate of the continuum SED of the disk. However,
the contribution of the upper layers to the SED or the role of the
deeper layers in millimeter images and emergent flux, would not be
consistent for this simple approach to settling. On the other hand,
taking into account the detailed physics of settling
\citep[e.g.][]{weidenschilling97, dullemond04} is complex and
simulations show that disks should be completely settled within
$\sim$10$^6$ years, in contradiction with observations, reflecting that
we are missing some processes that can keep some small grains in the
upper layers for longer timescales \citep[e.g.
turbulence;][]{dullemond05,birnstiel11}.

In our SED modeling we have adopted a different approach following
\citet{dalessio06} by parameterizing settling as a depletion of dust in
the upper layers, with a corresponding increment of the dust-to-gas mass
ratio near the midplane. The maximum grain sizes in the disk atmosphere
and interior are allowed to change, reflecting the possibility of grain
growth.  We can also vary the height in the disk that separates the
atmosphere from the interior as well as the degree of settling.  The
amount of settling is parameterized by
$\epsilon=\zeta_{atm}/\zeta_{std}$, (i.e.,  the dust-to-gas mass ratio
of the disk atmosphere divided by the standard value). The main point of
this approach is that the same grains that determine the height and
shape of the irradiation surface and the amount of intercepted external
flux, are the ones that are emitting in the mid-IR silicate bands, and
their emissivity and temperature distribution are consistent with their
properties. Also, the grains near the midplane which are responsible for
the mm emergent intensity have a dust-to-gas mass ratio related to the
properties of the atmospheric grains. The advantage of such an approach
is that, in principle, observations can be used to constrain the grains'
composition, size and spatial distribution, and this can be related to
models of the detailed dust evolution in disks. However, to really
fulfill this goal, we need observations that cover a wide range of
wavelengths with high resolution. Given our present observations, we
have chosen to adopt a radially constant $\epsilon$ and to assume that the
interior grains are concentrated very close to the midplane (at z
$\lesssim$ 0.1~H). These assumptions will not affect the mid-IR SED
\citep[][Qi et al. 2011]{dalessio06} and we avoid introducing new sets
of free parameters to the problem, retaining the important physical
properties of settling.

{\it Dust Grain Growth.} In this work we also change the maximum grain
size in the disk. The models assume spherical grains with a distribution
of $a^{-p}$ where $a$ is the grain radius between a$_{min}$ and
a$_{max}$ and p is 3.5 \citep{mathis77}. A mixture with a smaller
a$_{max}$ has a larger opacity at shorter wavelengths than a mixture with
a larger a$_{max}$. Since the height of the disk surface, $z_s$, is
defined by the region where $\tau_s \sim 1$ small grains will reach this
limit higher in the disk relative to big grains. Therefore, disks with a
small a$_{max}$ are more flared than disks with a large a$_{max}$ for
the same dust-to-gas mass ratio. One difference between increasing the
settling and increasing the grain size is that with settling, small
grains remain in the upper disk layers and so we still see silicate
emission while with grain growth in the disk atmosphere, the silicate
emission disappears since larger grains do not have this feature in
their opacity. In the walls and the outer disk,  $a_{min}$ is held fixed
at 0.005~{\micron} while $a_{max}$ is varied between 0.25~{\micron} and
10~{\micron} to achieve the best fit to the silicate emission features.
In the outer disk, there are two dust grain size distributions as
mentioned above. In the disk interior the maximum grain size is 1~mm
\citep{dalessio06}.  The maximum grain size of the disk atmosphere is
adjusted as noted earlier.

{\it Dust Composition.} The composition of dust used in the disk model
impacts the resulting SED and derived disk properties (see
\citet{espaillat10} for a discussion). We follow E11 and perform a
detailed dust composition fit for the silicates seen in the IRS spectra
including olivines, pyroxenes, forsterite, enstatite, and silica. We
list the derived silicate mass fractions in Tables~\ref{tab:silwall}
and~\ref{tab:silwall2} of the Appendix. In addition to silicates, we
also included organics, troilite, and water ice following
\citet{espaillat10} and E11. We note that only silicates exist at the
high temperatures at which the inner wall is located. In transitional
and pre-transitional objects where we include optically thin dust within
the hole, the silicate dust composition and abundances are listed in
Tables~\ref{tab:silthin} and~\ref{tab:thinprop} of the Appendix,
respectively.

\subsection{SED Modeling} \label{sec:modsed}

In this work we present the first detailed modeling of disks with IRS
spectra in NGC~2068 and IC~348.  We find that all the objects are
reasonably reproduced with transitional, pre-transitional, or full disk
models. 
It is not the goal of this paper to find a unique fit to the
SED.  To arrive at a unique fit, one would ideally have a finely sampled, multi-wavelength
SED as well as spatially resolved data at multiple wavelengths.  Finely sampled data on the time domain would also be necessary since the emission of TTS is known to be variable \citep[e.g.][]{espaillat11}.  Given that
this situation is not currently achievable, here we focus on finding a fit that is consistent with the
observations presented in this work.  
Our assumptions of the disk structure are an oversimplification.  Recent hydrodynamical
simulations show that the inner regions of transitional and pre-transitional disks
should be complex \citep{zhu11,dodson11}.  However, in the absence of data capable of confirming these simulations, we proceed with our simple model.
We discuss additional assumptions and
how they play into the degeneracies of our modeling in
Section~\ref{limitsmodels}.  We present details of the derived dust
composition in the Appendix.

\subsubsection{Results}

\begin{figure}
\epsscale{1}
\plotone{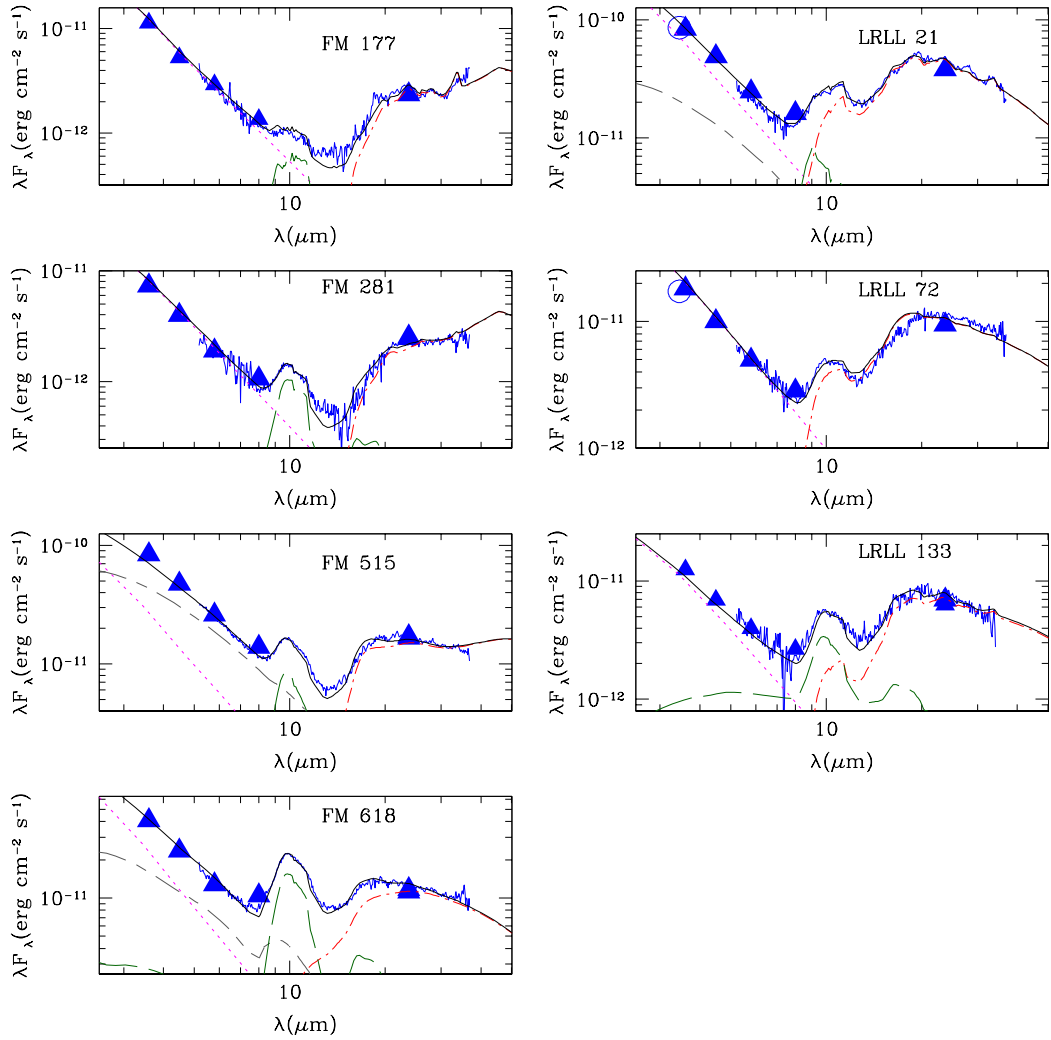}
\caption[]{IR SEDs and disk models (solid, black lines) of transitional 
and pre-transitional disks in our sample. Here we show only the
dereddened data from Figures~\ref{figsedtd} and~\ref{figsedptd}. Refer
to Section~\ref{sec:modsed} and Table~\ref{tab:disk} for model details.
Separate model components are the stellar photosphere (magenta dotted
line), the inner wall (gray short-long-dashed line), the outer wall (red
dot-short-dashed line), and the optically thin small dust located within
the inner disk (green long-dashed line). (A color version of this figure
is available in the online journal.)
}
\label{figmodeltd}
\end{figure}

\begin{figure}
\epsscale{1}
\plotone{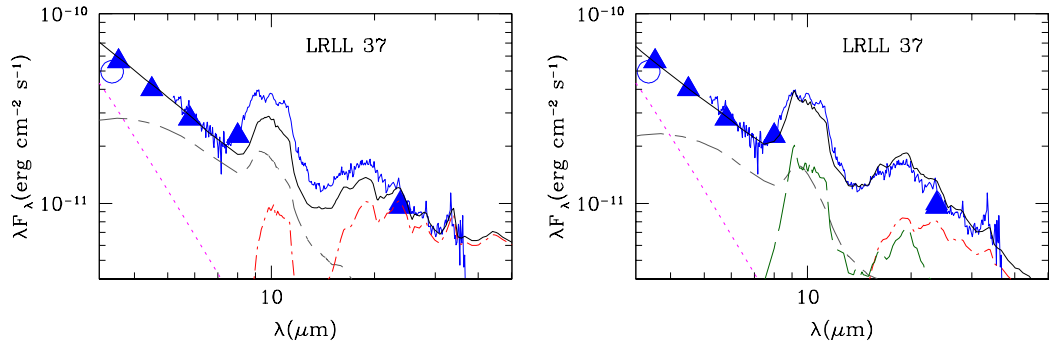}
\caption[]{Left: IR SED and full disk model fit of LRLL~37. We cannot successfully 
reproduce the 10~{\micron} and 20~{\micron} emission.  This disk model
has ${\epsilon}$=0.001, ${\alpha}$=0.0008, and a$_{max}$=0.25{\micron}.
Right: IR SED and pre-transitional disk model of LRLL~37.  We can fit
the strong silicate emission with a model that has a gap in the disk
with most of the silicate emission arising from small, optically thin
dust within the gap.  The color scheme, symbols, and lines are the same
as in Figure~\ref{figsedtd}. (A color version of this figure is
available in the online journal.)}
\label{figmodelic37}
\end{figure}

\begin{figure}
\epsscale{1}
\plotone{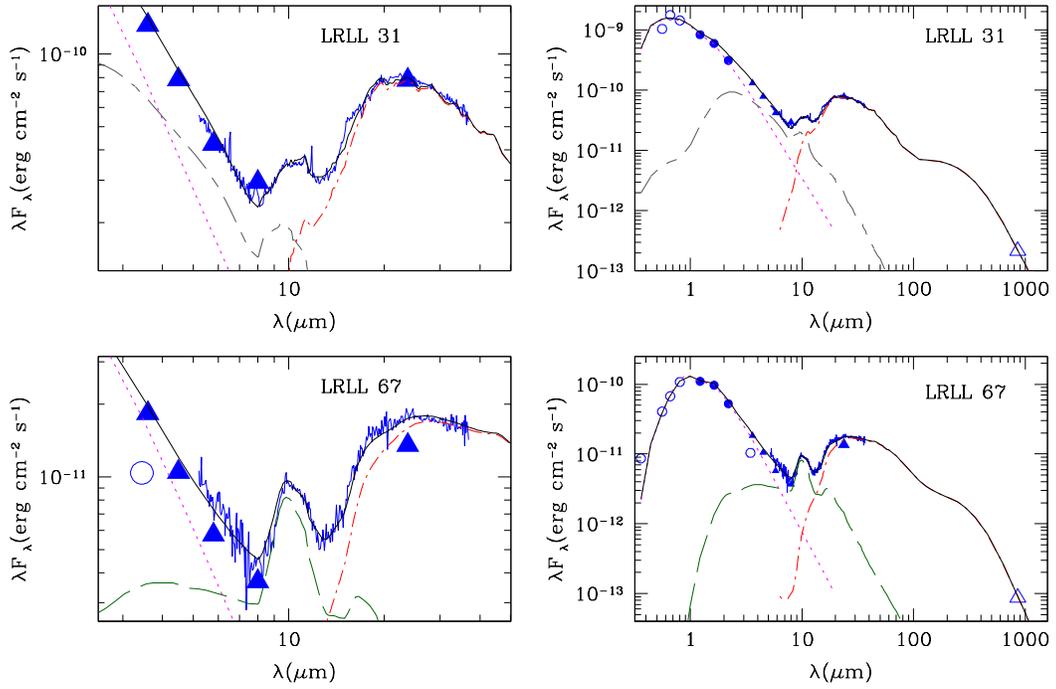}
\vskip -1.5in
\caption[]{Left: IR SEDs and disk models of the pre-transitional disk 
LRLL~31 and the transitional disk LRLL~67. Right: Broad-band SEDs and
disk models of LRLL~31 and LRLL~67 which include SMA millimeter data
(from this work). With millimeter data we can constrain the outer disk
and so we include both the outer wall and the disk behind it in the
model presented here (red dot-short-dash). We also show the contribution
to the SED from the inner wall of LRLL~31 (gray short-long-dash) and the
optically thin small dust located within the inner hole of LRLL~67
(green long-dash). Other symbols and lines are the same as used in
Figure~\ref{figsedtd}. (A color version of this figure is available in
the online journal.)
}
\label{figmodeltd2}
\end{figure}

We find a large range of hole and gap sizes for our transitional and pre-transitional
disks.  Our TD have holes spanning 4 to 49 AU   
(Figure~\ref{figmodeltd}, Table~\ref{tab:disk}).
Our pre-transitional disk targets have gaps ranging from 5--45~AU
 (Table~\ref{tab:disk} and
Figures~\ref{figmodeltd},~\ref{figmodelic37}, and~\ref{figmodeltd2}).
\footnote{As mentioned in Section~\ref{sec:diskmodel}, we do not include
an inner disk behind the inner wall.  \citet{espaillat10} find that the
inner wall dominates the NIR emission of PTD.  \citet{pott10}
confirm that the inner disk in PTD is small.  High resolution
imaging is needed to further constrain the gap sizes of the objects
presented in this work.}
  The three objects easily
identified by dips in the SED (FM~515, FM~618, LRLL~31) have gap sizes
of 11-45~AU. We note that here we classify LRLL~21 as a PTD even though
its NIR emission is weaker than the other PTD in our sample and it
resembles the emission expected from a TD.  \citet{flaherty12}
find that LRLL~21 has significant NIR emission in more recent IRS
observations, pointing to strong intrinsic variability in the inner disk
linked to changes in the inner wall \citep{espaillat11}.  Therefore,
here we classify  LRLL~21 as a PTD. Another PTD in our sample that is
not obvious based on its SED alone is LRLL~37 which has the smallest gap
size in our sample (5~AU; Figure~\ref{figmodelic37}). It is not possible
to fit the IRS data of LRLL~37 with a full disk model, even within the
uncertainties of the observations.  In particular, we could not fit the
strong 10~{\micron} silicate emission with our full disk model.  This
could be a sign that LRLL~37 is a pre-transitional disk with a small gap
that contains some small optically thin dust, reminiscent of RY~Tau
(E11) and we will return to this point in Section~\ref{limitsobs}. We
only have millimeter fluxes for two objects in our sample, LRLL~31 and
LRLL~67. For these disks we derive a disk mass of 0.06~{\msun} for each
(Figure~\ref{figmodeltd2}).  The best fitting $\epsilon$ and $\alpha$
for LRLL~31 are 0.001 and 0.005, respectively.  For LRLL~67,
$\epsilon$=0.001 and $\alpha$=4$\times$10$^{-5}$.

Most of the transitional and pre-transitional disks have small optically
thin dust within the inner 1~AU of the hole  or gap.  The exceptions are
the transitional disk LRLL~72, where we find the 10~{\micron} silicate
emission can be produced by the optically thin atmosphere of the outer
wall, and the pre-transitional disk LRLL~31, where the 10~{\micron}
silicate emission comes from the inner wall's atmosphere and optically
thin dust within the gap is not necessary to fit the observations.  The
mass of dust and sizes of the grains in this region are given in the
Appendix.

\begin{figure}
\epsscale{1}
\plotone{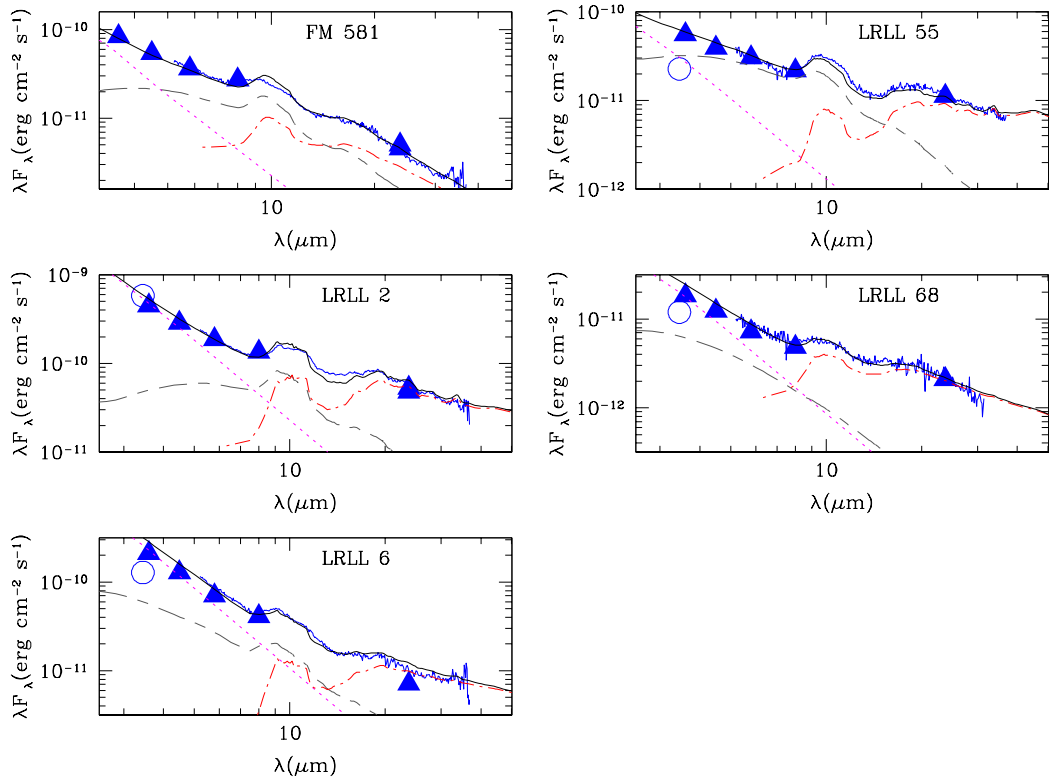}
\vskip -1.5in
\caption[]{IR SEDs and disk models of full disks in our sample.  
We can explain all of these disks with irradiated accretion disk models
which incorporate dust settling.  In the case of FM~581, we need to
truncate the outer disk radius to $<$1~AU in order to reproduce its very
steep slope. Refer to Section~\ref{sec:modsed} and Table~\ref{tab:disk}
for model details.  Here we show the contribution to the SED from the
stellar photosphere (magenta dotted line), the inner wall (gray
short-long-dash), and the outer disk (red dot-short-dash). (A color
version of this figure is available in the online journal.)
}
\label{figmodeled}
\end{figure}

For FM~581, LRLL~2, LRLL~6, LRLL~55, and LRLL~68 we can fit the SED
reasonably well using full disk models (Figure~\ref{figmodeled}). 
Unlike LRLL~31 and LRLL~67 above, we do not have millimeter detections
for FM~581, LRLL~2, LRLL~6, LRLL~55, and LRLL~68 and so we cannot
constrain the mass of the disk.  Therefore, the models presented here
are more uncertain, but we show them to illustrate that a disk model
with dust settling and no holes or gaps in the disk can reproduce the
observed SEDs. We do have an upper limit for the millimeter flux of
LRLL~68 and we use this object as an example to discuss the degeneracies
inherent in the modeling presented, mainly due to lack of millimeter
detections, in Section~\ref{appendix:ic68}.  To briefly summarize, disk
models with the same ${\epsilon}$-to-$\alpha$ ratio will produce very
similar emission in the IR but substantially different emission in the
millimeter.  Therefore, millimeter data is crucial to disentangle this
degeneracy and the disk parameters in this work should only be taken as
indicative of a model that can reproduce the observed SEDs.

With the above degeneracies in mind, we limited our parameter search and
set ${\epsilon}$=0.001, changing only $\alpha$ until we achieved a good
fit to the SED.  We could have also set $\alpha$ to a certain value and
fit for $\epsilon$ instead, however, as mentioned above and as discussed
in Section~\ref{appendix:ic68}, ${\epsilon}$$/$$\alpha$ is the most
relevant result.  For  LRLL~2, LRLL~55, and LRLL~68 we find
$\alpha$=0.06, 0.004, 0.006, respectively.  For LRLL~6 using an
${\epsilon}$ of 0.001 required $\alpha$$>$0.1 which would lead to a
viscous timescale shorter than the lifetime of the disk
\citep{hartmann98}, so instead we set ${\epsilon}$= 0.0001; the
best-fitting $\alpha$ in this case was 0.1. To fit the very steep
downward slope of FM~581, we needed to significantly truncate the outer
disk radius, down to 0.6~AU.\footnote{The SED of FM~581 resembles that
of the 5--10~AU binary SR 20 \citep{mcclure08}.  SR~20's disk is
outwardly truncated at 0.4~AU, too far to be attributable to the known
companion, and so this truncation would have to be due to an unseen
companion at $\sim$1--2~AU \citep{mcclure08}. Likewise, it seems that
the most viable mechanism to truncate the disk of FM~581 to such small
radii would also be a companion.  However, FM~581 still has a
substantial accretion rate.  Millimeter observations are necessary to
constrain the mass and size of this disk.}  We fit FM~581 with
$\epsilon$=0.001 and $\alpha$=0.00006.

\subsubsection{Model Degeneracies and Millimeter Constraints} \label{appendix:ic68}

Here we explore the degeneracies introduced into the modeling presented
in this paper due to the lack of millimeter data.  Rather than do this
for each object, we selected LRLL~68 for this test since it has an upper
limit to its millimeter flux at 860~{\micron} from the SMA observations
reported in this paper.

First, disk models (around the same star) with the same
$\epsilon$-to-$\alpha$ ratio will have SEDs with similar emission in the
IR. This is because disks with equal $\epsilon$$/$$\alpha$ have similar
disk surfaces. The disk surface is defined as the point in the upper
disk layers where the radial optical depth (which depends on the product
of the opacity and column density) to the stellar radiation reaches one.
$\epsilon$ determines the abundance of small dust (i.e. the opacity of
the upper disk layers) while $\alpha$ affects the surface density.
Therefore, in disks with equal $\epsilon$$/$$\alpha$ the same fraction
of stellar flux will be intercepted by the disk.  Since the IR is
dominated by the upper layers of the inner disk, their emergent
intensity in the IR will be similar.

While the IR emission is similar, the emission seen in the millimeter
will be different. A disk with an $\alpha-$viscosity has a mass surface
density given by $\Sigma \approx \dot{M} \Omega_k/\alpha <c_{s}>$, where
$\Omega_{k}$ is the Keplerian angular velocity and $<c_{s}>$ is the
sound speed. This implies that the disk mass, for a given disk radius
and similar outer disk temperature distribution,  would be proportional
to $\alpha$$^{-1}$.  Therefore, a disk with a smaller $\alpha$ will have
a larger disk mass.   In the millimeter, we are more sensitive to the
big grains in the midplane of the disk, where most of the disk's mass is
stored, and so disks with small $\alpha$ and a higher column density 
will have more millimeter emission. This means that models with similar
$\epsilon$cos$(i) {\mdot}/{\alpha}$ would have similar IR SEDs, but
different millimeter SEDs.  Millimeter data is necessary to disentangle
this degeneracy.

\begin{figure}
\epsscale{1}
\plotone{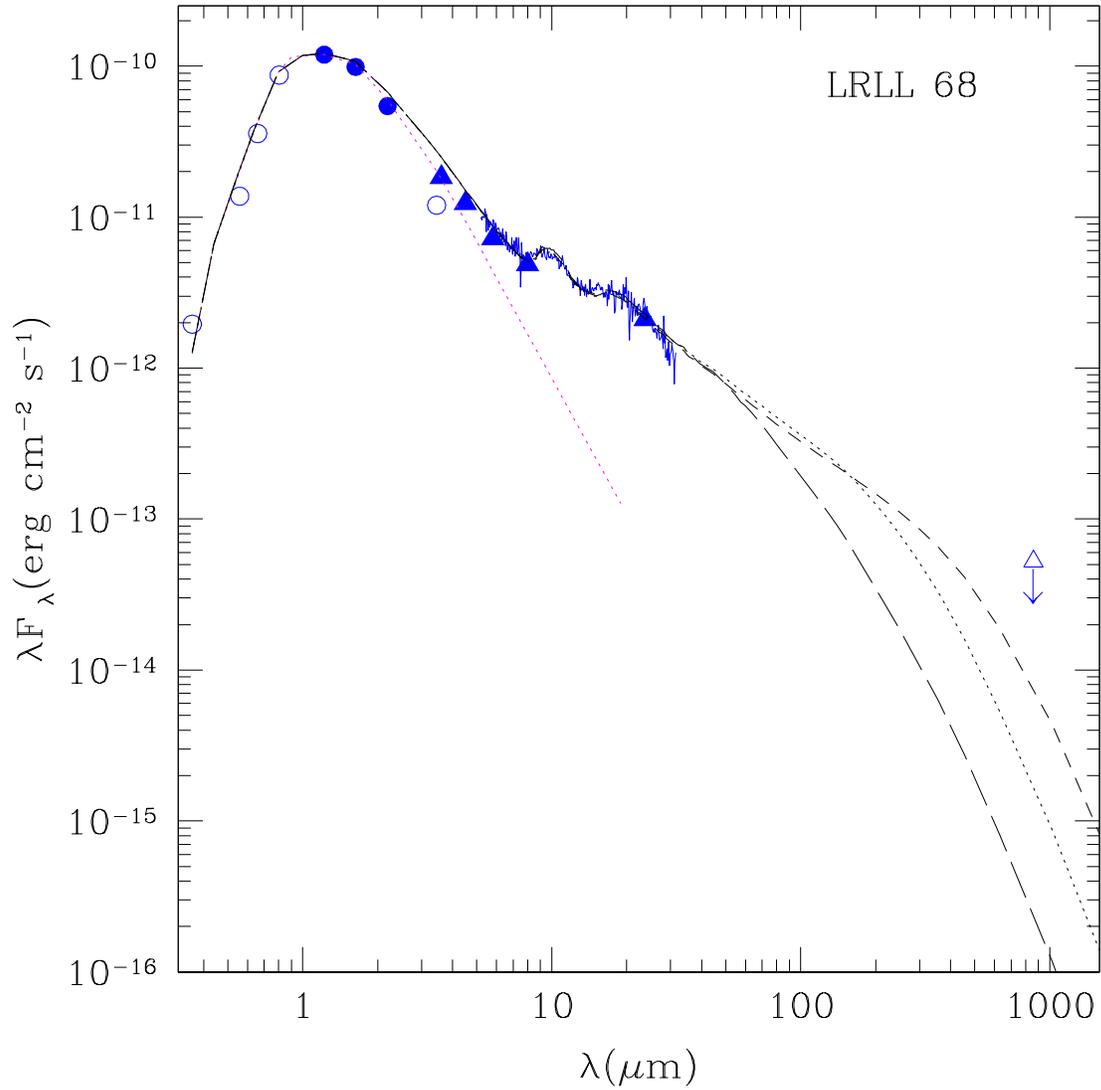}
\caption[]{Disk model fits to the SED of LRLL~68 with $\epsilon$$/$$\alpha$=6.  
Models with the same $\epsilon$-to-$\alpha$ ratio will have similar 
emission in the IR but substantially different emission in the millimeter.  
Here we show the following models from Table~\ref{tab:ic68}: Model 1 
($\epsilon$=0.0001; short-dashed line),  Model 2 ($\epsilon$=0.001; 
dotted line), and Model 3 ($\epsilon$=0.01; long-dashed line).  Note 
that each model falls below the millimeter upper limit presented in 
this work (open triangle).    This reflects the need for millimeter 
detections to constrain disk models.  
(A color version of this figure is available in the online journal.)
}
\label{figic68a}
\end{figure}

Because of the above, we can find a best-fit $\epsilon$$/$$\alpha$ to
the IR emission if we hold $i$ and ${\mdot}$ constant.  However, each
model will have different emission in the millimeter and since we have
no millimeter data, we cannot claim that a particular combination of
$\epsilon$ and $\alpha$ is better than another. For the models shown in
Figure~\ref{figic68a} (Models 1, 2, \& 3 in Table~\ref{tab:ic68}) we
hold the mass accretion rate, stellar parameters, inclination ($i$), and
outer disk radius $R_d$ fixed and change only $\epsilon$ and $\alpha$.
We set $\epsilon$ to 0.0001, 0.001, 0.01, and 0.1, the values given in
\citet{dalessio06}, and fit the SED by changing $\alpha$. We do not
discuss cases where $\alpha$$\geq$0.1 since these disks would have short
viscous timescales \citep{hartmann98}. For LRLL~68, we find that the
best-fitting $\epsilon$$/$$\alpha$ is 0.2.

\begin{figure}
\epsscale{1}
\plotone{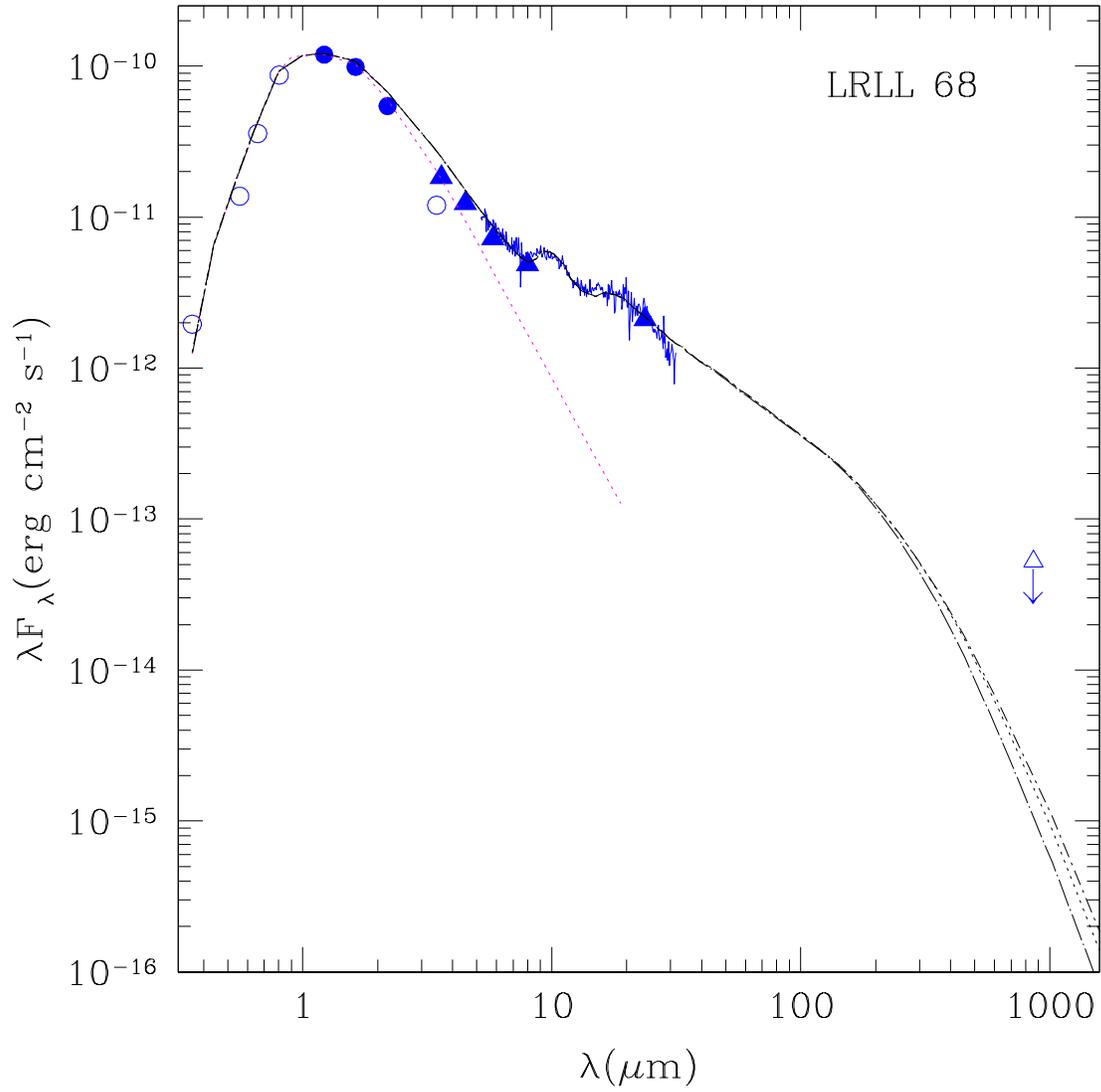}
\caption[]{Disk models of LRLL~68 with different R$_{d}$, keeping 
$\epsilon$ and $\alpha$ fixed.  Varying the outer radius of the disk will 
lead to changes in the disk mass, but small changes in the millimeter flux.  
Models shown are from Table~\ref{tab:ic68}:  Model 4 (R$_d$=300~AU; short- 
long-dashed line),   Model 2 (R$_d$=100~AU; dotted line), and Model 5 
(R$_d$=20~AU; dot- long-dashed line).
(A color version of this figure is available in the online journal.)
}
\label{figic68b}
\end{figure}

\begin{figure}
\epsscale{1}
\plotone{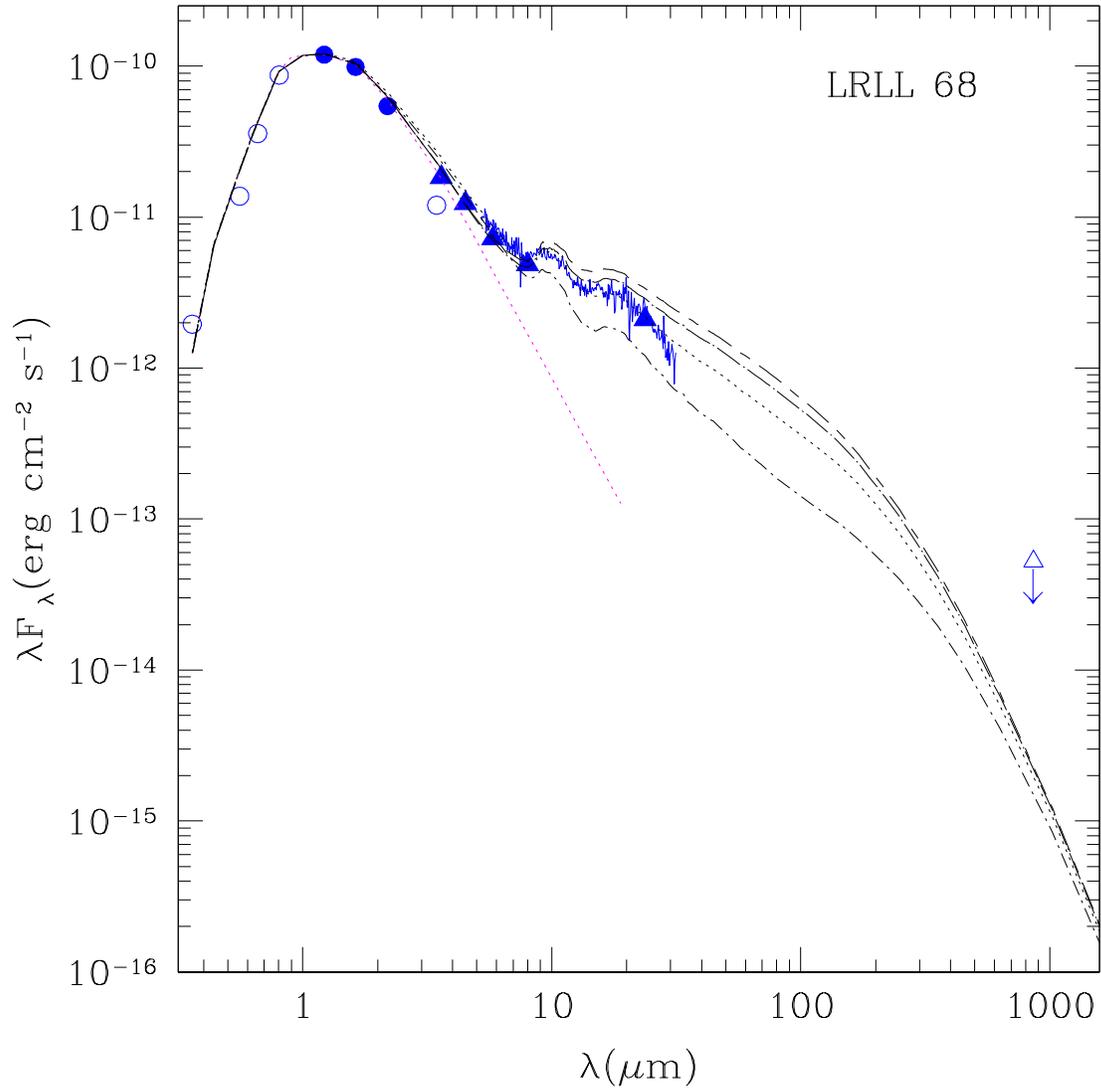}
\caption[]{Disk models of LRLL~68 with different $i$.  We find that 
changing the inclination of the disk, while keeping all other parameters 
fixed, will lead to substantial differences in the IR emission, but will 
not significantly affect the millimeter flux.  In order of increasing 
inclination (i.e. from nearly face-on to nearly edge-on) we show the 
following models from Table~\ref{tab:ic68}: Model 6 ($i$=20$^{\circ}$; 
short-long-dashed line), Model 7
($i$=40$^{\circ}$; dot-long-dashed line), Model 4 ($i$=60$^{\circ}$; 
dotted line), Model 8 ($i$=80$^{\circ}$; dot- short-dashed line).
(A color version of this figure is available in the online journal.)
}
\label{figic68c}
\end{figure}

We also look at how varying the outer disk radius and inclination affect
the simulated SED. Changing the radius changes the mass of the disk but
the millimeter emission does not change significantly
(Figure~\ref{figic68b}). This is because the mass depends on the disk
radius (M$_d$ = $\int^{R_{d}}_{R_{i}} \Sigma 2 \pi R dR$).  However, the
column density of the annuli contributing to the mm flux remains the
same. This highlights that disk sizes cannot be firmly constrained
without resolved imaging. We find that changing the inclination angle
while holding other parameters fixed changes the IR emission, but does
not significantly alter the millimeter emission (Figure~\ref{figic68c}).
The disk is mostly optically thin at millimeter wavelengths so we can
see through the disk at any inclination.  Therefore, the millimeter
emission does not change with inclination.   On the other hand, the IR
emission does change; as the inclination decreases, we see more of the
inner disk, which mainly dominates the IR emission.  We note that the
near-IR emission also depends on the shape of the wall. We assume the
wall is vertical, therefore it will not contribute to the SED at
0$^{\circ}$ or 90$^{\circ}$ and will produce the most emission at
60$^{\circ}$ \citep{dullemond01}.  If the wall is curved, we would
expect to see more emission at lower inclinations \citep{isella05}.

\section{Discussion} \label{sec:discuss}

\subsection{Limitations of Observations}\label{limitsobs}

As discussed by \citet{espaillat10}, the sizes of gaps and holes we can
detect in the disk are limited by broad-band SEDs. Given that the
majority of the  emission at $\sim$10 {\micron} in a typical disk traces
the dust within the inner 1 AU of the disk
\citep{dalessio06,espaillat09}, the {\it Spitzer} IRS instrument will be
most sensitive to clearings in which much of the dust located at radii
$<$1 AU has been removed. Because of this, IRS is more effective in
picking out disk {\it holes} where dust at small radii has been removed.
However, IRS cannot easily detect {\it gaps} whose inner boundary is
outside of $\sim$1 AU.  For example, a disk with a gap ranging from 5 --
10 AU will be difficult to distinguish from a full disk
\citep{espaillat10}. The gaps currently inferred solely from SEDs, which
have been modeled and imaged in the millimeter, are typically quite
large.  This reflects an observational bias towards picking out
pre-transitional disks with large gaps since their mid-infrared deficits
will be more obvious in {\it Spitzer} spectra. Smaller gaps will not
have as obvious of a deficit and will be difficult to detect. 
Broad-band colors from IRAC and MIPS have their limitations as well. 
They are useful for picking out TD, but it is difficult to distinguish
the NIR colors of a PTD from a full disk.

LRLL~37 may be a case of a disk with a small 5~AU gap.  The dip in the
SED is not obvious, but the strong silicate emission seen in this object
is reminiscent of RY Tau.  RY~Tau has a large cavity in its disk based
on millimeter imaging \citep{isella10} and its SED was modeled with a
gap of $\sim$20~AU (E11).  There are many other disks that exhibit
strong 10~{\micron} silicate emission \citep{furlan09} and perhaps this
is a hint pointing to small gaps in disks. That said, we cannot exclude
the presence of small gaps in what we have labeled full disks in
this paper, or for that matter any full disks in general.
High-resolution imaging is crucial to investigate this further and
current TD fractions should be taken as lower limits.

\subsection{Limitations of Models} \label{limitsmodels}

The underlying physical structure inferred from the SED is dependent on
the model one uses.  For example, there are five disks in Taurus with
decreasing emission at IR wavelengths, much like the disks studied in
this work, that have seemingly contrasting interpretations presented by
\citet{luhman10} and \citet{currie11}.  The main difference between the
two papers are the models adopted. \citet{currie11} use the model grid
presented in \citet{robitaille07}.  The authors account for the observed
SEDs by decreasing the mass of models with well-mixed gas and dust in
the disk (i.e. a disk with no settling) to the point where it becomes
entirely optically thin.  The \citet{luhman10} results are based on
synthetic colors from the model grid of \citet{espaillat09}. Those
models are the same in this work, where dust settling is incorporated. 
\citet{luhman10} can reproduce the observed colors with disks that have
dust settling.  To illustrate that a settled disk can reproduce the
broad-band SED as well, in the Appendix we model ZZ Tau, one of the five
disks in question. Therefore, it becomes clear that the different
interpretation between the two groups is biased by the models adopted.
The most one can say is that both a well-mixed low-mass disk and a
settled disk can reproduce the observations.  In the former case the
disk is optically thin to its own radiation at all radii; for the latter
case the innermost disk is still optically thick to its own radiation.

\citet{sicilia11} also independently find that they need to incorporate
settling to reproduce the SED of objects with decreasing IR emission. We
note that the implementation of settling in that work and our work are
very different. \citet{sicilia11} simulate settling in their modeling by
lowering the disk surface, but its shape remains the same.  In addition,
the dust-to-gas mass ratio is held fixed throughout the disk (i.e. there
is no dust depletion in the upper disk layers).  Since the surface is still flared and the
opacity remains the same, the disk is hotter than it would be if the
disk was geometrically flatter and the opacity was lower. Therefore,
such a disk will produce more emission and there will be an inherent
bias towards decreasing the disk mass in order to reproduce lower
observed fluxes. In the models used in this work we deplete the small
grains in the upper atmosphere of the disk and self-consistently
calculate the disk height and shape.  This is an iterative process given
that the height of the disk irradiation surface dictates how much energy
is captured by the disk and this depends on the opacity set by the dust
properties and the density of the disk, which in turn depends on the
temperature through the scale height. We note that it is still possible
to have a disk which is both settled and low-mass. Our point is that in
order to constrain the disk mass and avoid the degeneracies discussed in
Section~\ref{appendix:ic68}, millimeter data is necessary and the full
effects of dust settling need to be taken into account.

This leads to another issue that is quite model dependent:  the disk
mass.  Not only does the disk mass depend on the opacity one assumes, it
also depends on the surface density and temperature radial profiles. For
example, our mass determinations (i.e. E11) are consistently higher than
\citet{andrews05} since our opacities are $\sim$3 times lower, we assume
the outer disk radius is larger, and we use a self-consistent surface
density and temperature instead of power-law approximations.  
Therefore, masses obtained by different models cannot be meaningfully
compared.

\subsection{Disk Structure \& Semantics} \label{semantics}

There are many terms in the literature aiming to categorize SEDs of TTS
surrounded by disks.  At times this leads to confusion.  For example,
what some researchers call a transitional disk others would call an
anemic disk or an evolved disk or a homologously depleted disk. The
effect of not consistently applying the term ``transitional'' in the
literature is seen when looking at TD fractions in the literature.  Some
call ``transitional'' any objects whose SED does not resemble the median
SED of Taurus.  Others use the more restrictive definition of disks that
have holes. If we only consider objects with IR dips in their SEDs as
transitional, the TD fraction is lower; including objects with
decreasing emission at all wavelengths increases the reported
``transitional disk'' fraction from 20$\%$ to 70$\%$ at $\sim$10 Myr 
\citep{muzerolle10}.  This difference is not trivial and presents an
unclear picture when attempting to compare theoretical simulations of
planet formation that predict disk holes with a ``transitional
disk'' fraction that encompasses objects which do not have apparent
evidence for cleared regions in their disks.

Another related issue is encountered when trying to discern the disk
clearing timescale.  These timescales usually include transitional and pre-transitional
disks and evolved disks (which are optically thin), and are measured
with respect to full disks. Defining the boundaries between
transitional, pre-transitional, evolved, and full disks is crucial in order to
obtain an accurate estimate of the disk clearing timescale.  TD are
relatively easier to identify, whether looking at broad-band SEDs or
colors.  PTD are harder to tell apart from full disks based on colors
alone.  Evolved disks are also difficult to separate from full disks and
this is the main driving force behind the different disk clearing
timescales reported in the literature. \citet{currie11} find a
$\sim$1~Myr timescale for inner disk clearing, which is longer than the
$\sim$0.5~Myr timescale obtained by  \citet{luhman10}.  For the most
part, both groups are using the same data. The main difference is what
one considers a full disk versus evolved (also called homologously
depleted). As pointed out by \citet{hernandez10}, the reported fraction
of optically thin disks (i.e. evolved disks or homologously depleted
disks) is highly dependent on the cutoff.  Observationally,
\citet{currie11} use the lower quartile of Taurus.  Therefore, by
construction, 25$\%$ of disks in Taurus are evolved disks. 
\citet{luhman10} use a gap in the IR color-color diagrams which leads to
fewer objects in this phase.

Here we focus on the physical structure that could be underlying the
observed SEDs by using physical models to motivate our interpretation.
We find that it is possible to group objects based on their observed
SEDs and find a general model-based interpretation to fit objects within
a group.  In our work we have cases of disks with holes and gaps and
full disks (see Section~\ref{sec:modsed}).
We emphasize, as noted previously in Section~\ref{limitsobs},
that possibly {\it all} disks have gaps that cannot be detected in SEDs.
However, our goal here is to discern when one {\it can} identify a hole
or gap in a disk based on its SED.

 In essence, all disks around
TTS are ``in transition'' as they are all evolving in one way or
another.  The expectation is that all TTS with disks will eventually
become diskless stars.  However, they are not all necessarily going down
the same evolutionary path. Here we suggest that the term
``transitional'' be used for a disk which appears to be undergoing a
radical disturbance in the radial structure of its inner disk (i.e., a hole or gap). While
above we point out the inherent deficiencies in using the evolutionary
term ``transitional'' to define a disk, introducing new classification
schemes to the literature is not warranted given the limitations of
currently available data.  

One motivation for separating disks with holes and gaps from full disks
is that it is not obvious these disks are undergoing the same type of
disk clearing. Many researchers have suggested that the holes and gaps
in disks observed to date are due to planets \citep[see discussion
in][]{espaillat10}. Simulations have shown that newly forming planets
will clear regions of the disk through accretion and tidal disturbances
\citep{goldreich80, ward88, rice03, paardekooper04, quillen04,
varniere06, zhu11}. It is less clear how a disk with weak emission at
all wavelengths could be related to disk clearing caused by a planet. As
pointed out by \citet{cieza10} and \citet{currie11}, disks with weak MIR
emission could instead be the result of another mechanism that may
have a different rate of evolution (e.g. photoevaporation).
Alternatively, full disks with SEDs such as those in this paper could
simply be the tail end of continuous distribution of full disks.   This
could be related to a large spread in disk properties (${\mdot}$, M$_d$,
dust composition) in a given population as well as a distribution in the
initial conditions. Another possibility, that we cannot fully test in
this paper due to a lack of millimeter detections, is that the full
disks in our sample have experienced a greater degree of dust settling
than other full disks.  In this case, we would see more of these disks
in older regions since settling is expected to increase with age.

\subsection{Mass Accretion Rates of Transitional and Pre-transitional Disks} \label{tdmdot}

\citet{najita07a} showed that the mass accretion rates of transitional
disks in Taurus tend to be lower than those of full disks in the same
region.  If planets are the clearing agent in transitional disks, then
lower mass accretion rates are expected onto the star since a giant
planet that opens a gap in the disk will intercept and accrete material
from the outer disk \citep{lubow06}. To explore this further we compared
the distribution of mass accretion rates of full disks and
transitional and pre-transitional disks in Taurus, Chamaeleon, and NGC~2068
(Figure~\ref{fighist}; see the Appendix for details).  We note that
\citet{najita07a} used a broader definition of ``transitional disk''
which included all objects with less emission than the median SED of
Taurus.  As discussed in Section~\ref{semantics}, the link between these
disks and planet formation is less clear.  Therefore, here we use our
more restrictive definition of transitional and pre-transitional disks which
includes only objects with holes and gaps.  We find that the mass
accretion rates of transitional and pre-transitional disks tend to be about 5 times
lower than the full disks in these three regions.  The median mass
accretion rate for the full disks is 1.3$\times$10$^{-8}$ ${\msun}$
yr$^{-1}$ and for the transitional and pre-transitional disks it is
3.1$\times$10$^{-9}$ ${\msun}$ yr$^{-1}$.  A Kolmogorov-Smirnov (KS)
test indicates that the full disk and transitional and pre-transitional disk mass
accretion rate samples are not drawn from the same distribution (the
KS-probability is 0.02).

While the mass accretion rates of transitional and pre-transitional disks are overall
lower than those of full disks, they are still too high to be compatible
with current models of disk clearing by planets.  This is especially
seen in the cases of transitional and pre-transitional disks with higher mass accretion
rates and large gaps and holes.  \citet{zhu11} find that multiple
planets are needed to open these large clearings in the dust
distribution.  However, more planets in the disk should lead to lower
mass accretion rates onto the star than those observed.  Our results
suggest that possible planets in transitional and pre-transitional disks could be
lowering the mass accretion rate onto the star somewhat, but that there
is another mechanism taking effect that we have not accounted for,
possibly dust evolution as proposed by  \citet{zhu11}.  More simulations
of disk clearing by planets are needed to reconcile the large gap sizes
and mass accretion rates currently observed.

We also see that the transitional disks tend to have have lower mass
accretion rates than the pre-transitional disks in our sample, by a
factor of $\sim$10. The median mass accretion rate for our five
transitional disks is 9.7$\times$10$^{-10}$ ${\msun}$ yr$^{-1}$ while
the median for the ten pre-transitional disks in the sample is
8.8$\times$10$^{-9}$ ${\msun}$ yr$^{-1}$.  (More observations of PTD and
TD are needed to expand the sample size and confirm this result given
that the KS-probability that the samples are drawn from the same
distribution is 0.27.)  
Given that the evolution and relationship between TD and PTD
is not currently completely understood, the underlying reason for this 
apparent discrepancy in mass accretion rates is not obvious.
One can speculate that the difference is due
to the same mechanism clearing the holes and gaps in these disks.
In the case of planet formation,
\citet{zhu11} find that the mass accretion rate
onto the star will decrease with time as planets grow in the disk.   The
difference in mass accretion rate between PTD and TD could then possibly indicate that
pre-transitional disks are in the early stages of planet formation while
transitional disks are in the later stages.  However, refinement of planet
forming simulations is needed to study this
further given the complex structure expected in the inner disk region.

\section{Summary} \label{sed:sum}

Here we modeled the broad-band SEDs of 15 disks in NGC~2068 and IC~348. 
We presented IRS spectra for all our targets as well as mass accretion
rates estimated with U-band photometry obtained at the MDM Observatory
and H$_{\alpha}$ profiles from the MIKE spectrograph on the Magellan
telescope.  We also presented SMA millimeter data for some of our
sources in IC~348.

The observed SEDs of the objects in our sample are diverse, yet can be
separated into three groups. Some of our targets have dips in both their
NIR and MIR emission, some have dips in only their MIR emission, and
some have decreasing emission at all IRS wavelengths.  We modeled the first
group as transitional disks (i.e. objects with holes in their disk's dust
distribution), the second group as pre-transitional disks (i.e. objects
with gaps in their disk's dust distribution), and the last group as full disks
(i.e. objects with no cleared regions in their disks). We found that millimeter data are crucial in
breaking model degeneracies between the amount of dust settling in the
disk and the disk's mass.

We discussed the limitations of the observations, namely that we
currently do not have high enough resolution to discern very small gaps
in disks, and the limitations of disk models, especially with respect to
simulating the effects of dust settling and determining masses. We
pointed out that much of the disagreement in the literature over
reported transitional disk frequencies and disk clearing timescales is
mainly due to inconsistent application of the term ``transitional'' in
the literature.  We suggested that only objects showing evidence of an
abrupt change in their radial disk structure be referred to as
``transitional.''   Specifically, here we use ``transitional disk'' when
referring to disks with holes and ``pre-transitional disk'' for disks with gaps.
Finally, we compared the mass accretion rates of
transitional and pre-transitional disks to full disks in Taurus, Chamaeleon, and
NGC~2068 and find that PTD and TD have lower accretion rates overall. 
We also find that the TD have lower mass accretion rates than PTD, but
due to our small sample more objects are needed to confirm this.

Significant progress will be made in the near future on the issues
raised in this paper.  {\it Herschel} SPIRE will provide us with a
large, consistent sample of sub-millimeter fluxes to help break model
degeneracies. With the high resolution of {\it ALMA} we can soon test if
the above classifications used in this paper hold and modify them if
necessary.

 \acknowledgments{ We thank Lee Hartmann for helpful discussions and the referee for 
 a careful review of the manuscript.
 C.~E.~was supported by the National Science Foundation under Award No. 0901947.  
P.~D.~acknowledges a grant from PAPIIT-DGAPA UNAM.
N.~C.~acknowledges support from NASA Origins Grant NNX08AFM~5154G. 
}

\bibliographystyle{apjv2}

\appendix

\section{Dust Composition of Sample}
We performed fitting of the silicate emission features visible in the
IRS spectra and derived the mass fraction of amorphous and crystalline
silicates in the disk (Tables~\ref{tab:silwall},~\ref{tab:silwall2}
and~\ref{tab:silthin}). See E11 for a discussion of the degeneracies
inherent in deriving the dust composition. The results here should be
taken as representative of a dust composition that can reasonably
explain the observed silicate features in the SED. We leave it to future
work to further constrain the mass fractions of silicates in these
disks.

For the inner wall of our pre-transitional objects, we adopted a
silicate composition consisting solely of amorphous olivines.  This is
because the inner wall does not produce significant 10~{\micron}
silicate emission in most of the objects in this study and so we have no
way to distinguish between pyroxene and olivine silicates in the inner
wall.  The exception is LRLL~31 where we need 60$\%$ amorphous olivine
and 40$\%$ forsterite in the inner wall in order to fit the 10~{\micron}
silicate emission feature.  For transitional and pre-transitional disks
we changed the silicate composition in the optically thin dust region
and outer wall to fit the SED (Tables~\ref{tab:silwall2}). For the full
disks, we changed the silicate composition in the inner wall and disk
(Table~\ref{tab:silwall} and~\ref{tab:silthin}). The silicate
composition was not allowed to vary between the inner wall and disk in
the full disk models.

\section{Comments on Fitting ZZ Tau with a Settled, Irradiated Accretion Disk Model}

Here we present modeling of the SED of ZZ Tau using
the disk model of \citet{dalessio06} discussed in
Section~\ref{sec:diskmodel}.  Stellar parameters used in the disk model
(T$_*$=3470~K; L$_*$=0.75~${\lsun}$; M$_*$=0.35~{\msun};
R$_*$=2.4~{\rsun}) were derived in the same manner as other objects in
this work (see Section~\ref{sec:starprop}) using a spectral type of M3
adopted from \citet{kh95} and a visual extinction (A$_V$) of 0.98 taken
from \citet{furlan06}. We adopt a mass accretion rate of
1.3$\times$10$^{-9}$ ${\msun}$ yr$^{-1}$ from \citet{white01}. We note
that this is slightly higher than the mass accretion rate measured from
U-band photometry (9$\times$10$^{-10}$ ${\msun}$ yr$^{-1}$).

In Figure~\ref{figzztau} we present two models with different outer
radii.  In one model we use an outer disk radius of 100~AU for
comparison with previous modeling performed by \citet{currie11}. The
best-fit parameters are $\epsilon$=0.001 and $\alpha$=0.02 and this disk
has a mass of 5$\times$10$^{-4}$~{\msun}.  We also explored disks with
other $\epsilon$ values. A disk with a higher $\epsilon$ of 0.01 needs
an $\alpha$ of 0.2 to the fit the SED, but this $\alpha$ results in a
viscous timescale shorter than the lifetime of the disk
\citep{hartmann98}.  A disk with $\epsilon$=0.0001 and $\alpha$=0.002
with a higher mass of 0.005~${\msun}$ is excluded by the millimeter
upper limits.  Here we use amorphous silicates to fit the disk with
a$_{max}$=10~{\micron}. Crystalline silicate features are evident in the
IRS spectrum \citep{sargent09}, but we leave a detailed fit to McClure et al. (in
preparation). In Figure~\ref{figzztautau}, we show that ZZ Tau is
optically thick to its own radiation ($\tau_{Ross}$$>$1) out to about
$\sim$1~AU in the disk.  Over 80$\%$ of the emission seen at
40~{\micron} is from within these radii \citep[see Figure 2.16
in][]{espaillat09}.  Therefore, the optically thick part of the disk of
ZZ Tau dominates the emission seen in the IRS spectrum.

We also present a model with an outer radius of 3~AU in
Figure~\ref{figzztau}. This is because ZZ Tau is a close binary with a
separation of 0.06$^\prime$$^\prime$ \citep{schaefer06}, which
corresponds to 8~AU at the distance of Taurus (140~pc).  Therefore, the
IRS spectrum presented here includes both objects.  (ZZ~Tau~IRS, which
is 36$^\prime$$^\prime$away, did not enter the IRS slit and could be
a wide companion \citep{furlan11}.)   If a circumbinary disk is present,
its inner edge would be located at $\sim$16~AU according to expectations
of dynamical clearing by companions \citep{artymowicz94}.  However, we
detect NIR blackbody emission which indicates that instead we are seeing a
circumprimary disk.  In this case, the outer edge of the disk would be
truncated to $\sim$3~AU \citep{artymowicz94} and have a mass of
2$\times$10$^{-5}$~{\msun}. As discussed in Section~\ref{appendix:ic68}
and shown in Figure~\ref{figzztautau}, since the IRS emission is
dominated by the inner AU of the disk, changing the outer radius of the
disk does not significantly alter the IR emission.

\section{Comments on the Distribution of Mass Accretion Rates}

\begin{figure}
\epsscale{1}
\plotone{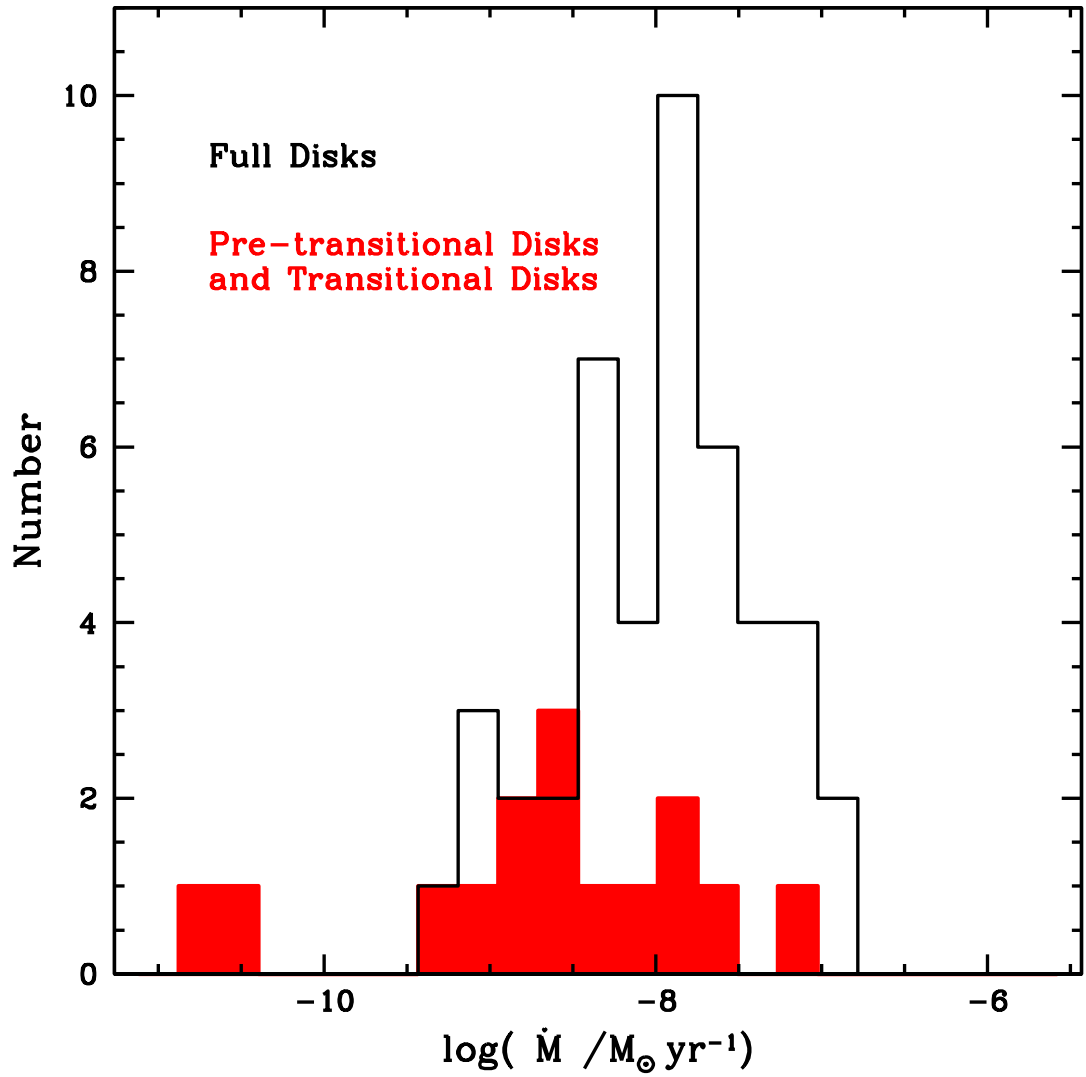}
\caption[]{Distribution of mass accretion rates for Taurus, Chamaeleon, 
and NGC~2068.  We separate full disks (white area) and transitional and pre-transitional 
disks (shaded area); overall, transitional and pre-transitional disks have lower mass 
accretion rates than full disks.  (A color version of this figure is 
available in the online journal.)
}
\label{fighist}
\end{figure}

\begin{figure}
\epsscale{1}
\plotone{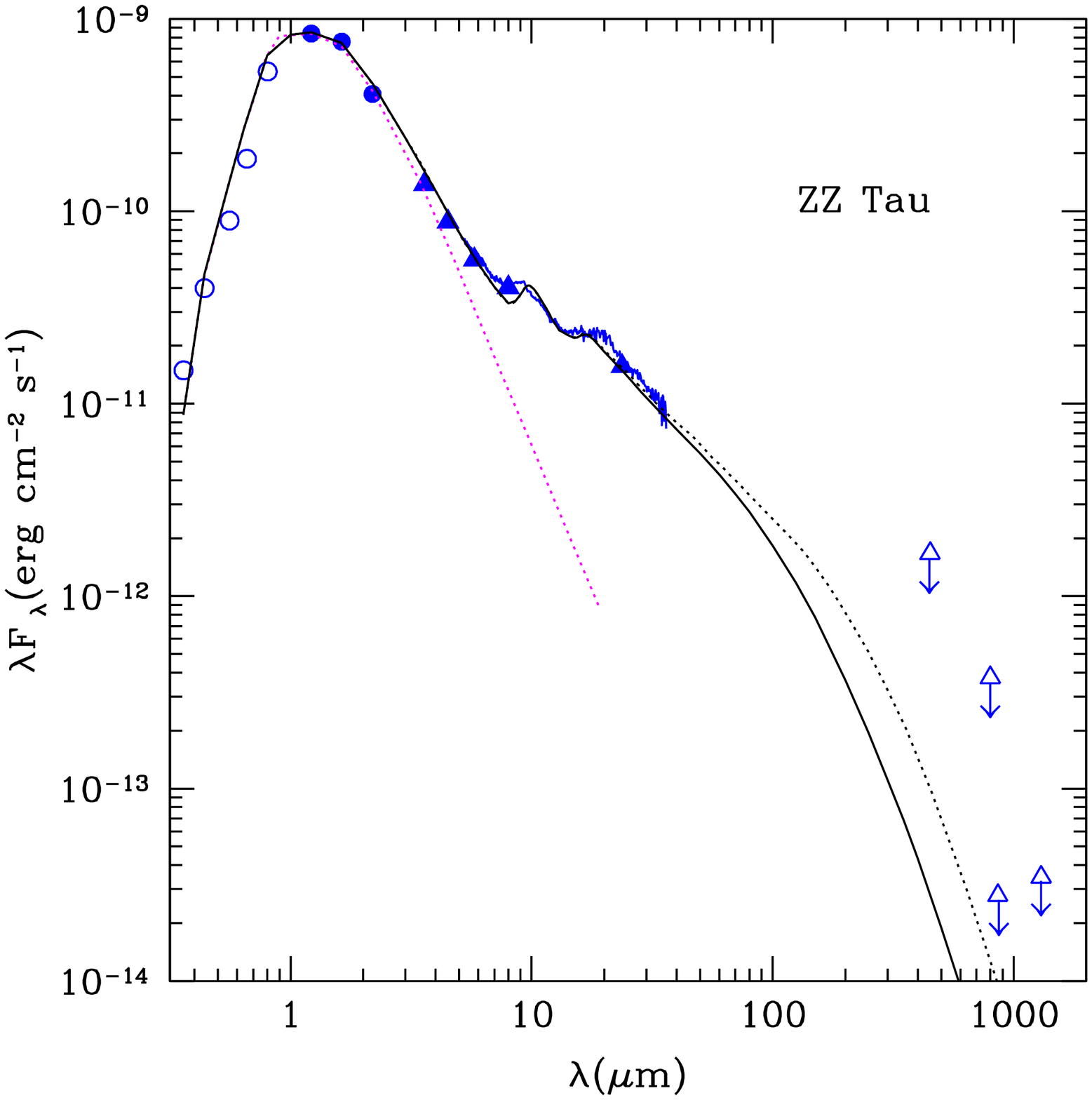}
\caption[]{Models of ZZ Tau using an irradiated accretion disk with dust settling.   We show a disk with
amorphous silicates and an outer radius of 3~AU (solid) and 100~AU (broken). The color scheme,
symbols, and lines are the same as in
Figures~\ref{figsedtd}~and~\ref{figmodeled}.  U-,B-, V-, R-, and I-
photometry are from \citet{kh95}; J-, H-, and K-band data are from
\citet{skrutskie06}; {\it Spitzer} IRAC and MIPS data are from
\citet{luhman10} and the IRS spectrum is from \citet{furlan06}.  All
millimeter upper limits are taken from \citet{jensen94} and
\citet{andrews05}. (A color version of this figure is available in the
online journal.)
}
\label{figzztau}
\end{figure}

\begin{figure}
\epsscale{1}
\plotone{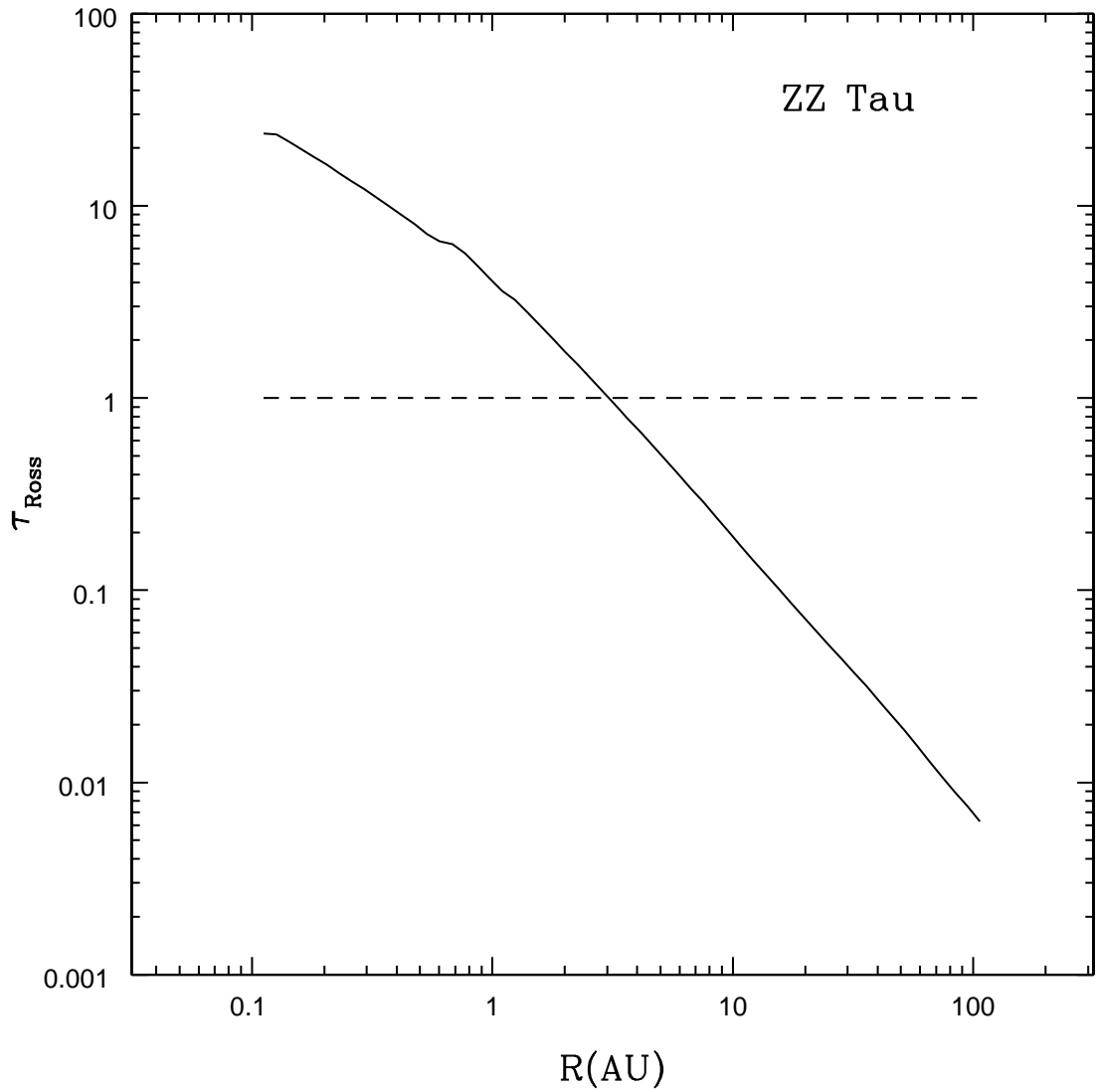}
\caption[]{Rosseland mean optical depth of ZZ Tau for the model shown in 
Figure~\ref{figzztau}.  ZZ~Tau is optically thick to its own radiation
($\tau_{Ross}$$>$1) out to $\sim$1~AU in the  disk.  It is these
innermost disk radii that dominate the emission seen in the IRS
spectrum.
}
\label{figzztautau}
\end{figure}

When plotting the distribution of mass accretion rates of full disks and
transitional and pre-transitional disks in Taurus, Chamaeleon, and NGC~2068
(Figure~\ref{fighist}) we restricted ourselves to transitional and pre-transitional
disks whose SEDs have been modeled. Mass accretion rates for
transitional and pre-transitional disks are taken from \citet{espaillat11} for Taurus
and Chamaeleon and from this work for NGC~2068.  The mass accretion
rates for full disks in Taurus were taken from \citet{najita07a}.  Since
here we use a different definition of ``transitional disk'' than used in
that work, we use the mass accretion rates for objects that do not
overlap with what we label a pre-transitional or transitional disk.  For Chamaeleon,
mass accretion rates for full disks are from \citet{hartmann98} and for
NGC~2068 mass accretion rates for full disks are from
\citet{flaherty08}.   We do not include objects which have upper limits
for their mass accretion rates or those that are known to be binaries. 
We also excluded IC~348 in this analysis since, to the best of our
knowledge, there are no mass accretion rates in the literature for the
full disks in this region.  In addition, IC~348 is older than Taurus,
Chamaeleon, and NGC~2068 which may bias the results given that mass
accretion rates are known to decrease with age
\citep[e.g.][]{calvet05b}.  In total we have 45 full disks and 15
transitional and pre-transitional disks.

We note that the majority of these mass accretion rates are derived using
U-band photometry and the relation in \citet{gullbring98}.  The exceptions are objects
in NGC~2068.  The mass accretion rates for transitional and pre-transitional
disks in NGC~2068 are taken from this work and the mass accretion rates
for the full disks are adopted from \citep{flaherty08}.  In Section~\ref{sec:mdot},
we discuss the derivation methods used in both works.  In short, the typical error (a factor of 2--3) inherent
to mass accretion rate estimation methods should not lead to systematic differences
between different samples.

\clearpage

\begin{deluxetable}{llll}
\tabletypesize{\scriptsize}
\tablewidth{0pt}
\tablecaption{Target Sample \label{tab:targetlog}}
\startdata
\hline
\hline
Name &  Region & RA & DEC \\
\hline
FM~177 		&  NGC 2068 	& 05h45m42s & --00d12m05s   \\  
FM~281 		&  NGC 2068 	& 05h45m53s & --00d13m25s    \\  
FM~515 		&  NGC 2068 	& 05h46m12s  & +00d32m26s   \\ 
FM~581 		&  NGC 2068 	& 05h46m19s & --00d05m38s  \\
FM~618 		&  NGC 2068 	& 05h46m23s & --00d08m53s    \\ 
LRLL~2		&	  IC 348			& 03h44m35s & +32d10m04s   \\ 
LRLL~6		&	  IC 348			& 03h44m37s & +32d06m45s  \\ 
LRLL~21		&	  IC 348			& 03h44m56s & +32d09m15s \\ 
LRLL~31		&	  IC 348			& 03h44m18s & +32d04m57s  \\ 
LRLL~37 		& IC 348			& 03h44m38s & +32d03m29s  \\  
LRLL~55 		&	 IC 348			&  03h44m31s & +32d00m14s \\  
LRLL~67		&	IC 348			& 03h43m45s & +32d08m17s   \\ 
LRLL~68 		&	IC 348			& 03h44m29s & +31d59m54s \\  
LRLL~72 		&	IC 348			& 03h44m23s & +32d01m53s  \\  
LRLL~133 		& IC 348			& 03h44m42s & +32d12m02s  
\enddata
\tablecomments{Target ID's are taken from \citet{flaherty08} and  \citet{luhman03} for targets in NGC~2068 and IC~348, respectively.
}
\end{deluxetable}

\begin{deluxetable}{lcccc}
\tabletypesize{\scriptsize}
\tablewidth{0pt}
\tablecaption{IC~348 Optical Photometry \label{tab:phot}}
\startdata
\hline
\hline
Target & V  & U--V & V--R  &  V--I   \\
 \hline
LRLL~2		& sat.  & sat. & sat. & sat.  \\
LRLL~6		& sat.  & sat. & sat. & sat.  \\
LRLL~21		& 15.68$\pm$0.03  & 3.51$\pm$0.06 & 1.35$\pm$0.08 & sat. \\
LRLL~31		& 19.30$\pm$0.08  & $<$0.8 & 2.00$\pm$0.09 & 3.89$\pm$0.09  \\
LRLL~37		& 15.85$\pm$0.04  & 2.68$\pm$0.25 & 1.24$\pm$0.05 & 1.19$\pm$0.06 \\ 
LRLL~55		& 21.68$\pm$0.33  & $<$--1.58 & 2.03$\pm$0.22 & 4.19$\pm$0.38  \\
LRLL~67		& 16.23$\pm$0.02  & 2.04$\pm$0.11 & 1.16$\pm$0.04 & 2.50$\pm$0.04  \\
LRLL~68		& 17.49$\pm$0.03  & 2.53$\pm$0.31 & 1.64$\pm$0.03 & 3.46$\pm$0.03 \\
LRLL~72		& 17.55$\pm$0.03  & 2.38$\pm$0.20 & 1.62$\pm$0.03 & 3.32$\pm$0.03  \\
LRLL~133		& 20.00$\pm$0.30  & $<$0.1 & 1.50$\pm$0.50 & 3.70$\pm$0.30  
\enddata
\tablecomments{We use ``sat.'' to refer to observations that were saturated and note upper limits for bands in which
sources were not detected.
 } 
\end{deluxetable}

\begin{deluxetable}{llllllllll}
\tabletypesize{\scriptsize}
\tablewidth{0pt}
\tablecaption{Source Properties \label{tab:prop}}
\startdata
\hline
\hline
Target & A$_V$ & Spectral  & T$_{*}$ &  L$_{*}$          &M$_{*}$         & R$_{*}$          &  $\mdot$                       		             & L$_{acc}$ & $\mdot$ \\
             &               &Type        &(K)           & (M$_{\sun}$) & (M$_{\sun}$) & (R$_{\sun}$) &  ( 10$^{-8}$ M$_{\sun}$ yr$^{-1}$) & (L$_{\sun}$) & Source  \\
\hline
FM~177 	& 1.6 & K4 	& 4590 & 1.0 & 1.2 	& 1.5 & 0.004 	& 0.0009 	& H$\alpha$   \\  
FM~281 	& 2.0 & M1 	& 3720 & 0.4 & 0.5 	& 1.6 & 0.002 	& 0.0002 	& H$\alpha$ \\  
FM~515 	& 1.6 & K2 	& 4900 & 2.5 & 1.5 	& 2.2 & 3.10 & 0.68			&  H$\alpha$ \\ 
FM~581 	& 4.1 & K4 	& 4590 & 4.1 & 1.6 	& 3.1 & 2.57 & 0.40 &  H$\alpha$ \\
FM~618 	& 2.9 & K1 	& 5080 & 2.2 & 1.5 	& 1.9	 & 1.21 & 0.29 &  H$\alpha$ \\ 
LRLL~2	& 3.8 & A2 	& 8970 & 57.1 & 2.8 & 3.1 & --		& --	& -- \\ 
LRLL~6	& 3.9 & G3 	& 5830 & 16.6 & 2.4 	& 4.0 & --		&--		& -- \\ 
LRLL~21	& 4.7 & K0 	& 5250 & 3.8 & 1.6 	& 2.4 & 0.20 	& 0.04	 	& H$\alpha$ \\ 
LRLL~31	& 8.6 & G6 	& 5700 & 5.0 & 1.6 	& 2.3 & 1.4 	& 0.3		& F11 \\ 
LRLL~37 	& 2.8 & K6 	& 4205 & 1.3 & 0.9 	& 2.2 & 0.13 	& 0.02		&  U-band \\  
LRLL~55 	& 8.5 & M0.5 	& 3850 & 1.0 & 0.6 	& 2.2 & --		&--		&  --  \\  
LRLL~67	& 2.0 & M0.75 	& 3720 & 0.5 & 0.5 	& 1.8 & 0.01 	& 0.001		&  H$\alpha$\\ 
LRLL~68 	& 2.1 & M3.5	 &3470 & 0.5 & 0.3 	& 2.0 & 0.04 	& 0.002		&  U-band \\  
LRLL~72 	& 3.0 & M2.5 	& 3580 & 0.7 & 0.4 	& 2.1 & $<$0.0003	& $<$0.00001	&  H$\alpha$ \\  
LRLL~133 	 & 3.6 & M5 	& 3240 & 0.2 & 0.2	& 1.5 & $<$0.8		& $<$0.78	&   U-band   
\enddata
\tablecomments{Spectral types for objects in NGC ~2068 and IC~348 are adopted from \citet{flaherty08} and \citet{luhman03}, respectively, except in the case of LRLL~31 where we adopt the spectral type of \citet{flaherty11}.  T$_{*}$  is taken from \citet{kh95}, based on the adopted spectral type.  L$_{*}$, M$_{*}$ and R$_{*}$ are calculated in this work.  A$_{V}$ measurements for most of the objects are from this work except for LRLL~2 and LRLL~6 where this value is adopted from \citet{luhman03}. The last column lists the method with which our mass accretion rates were calculated: H$\alpha$ and U-band are from this work; F11 is from \citet{flaherty11}.  For sources where mass accretion rate estimates are not listed here, we adopt 1$\times$10$^{-8}$ M$_{\sun}$ yr$^{-1}$.
 } 
\end{deluxetable}

\begin{deluxetable}{lccclccccccc}
\tabletypesize{\scriptsize}
\tablewidth{0pt}
\tablecaption{Wall Properties of Sample\label{tab:disk}}
\startdata
\hline
\hline\multicolumn{1}{c}{Target} & \multicolumn{1}{c}{Disk}& \multicolumn{1}{c}{}& \multicolumn{4}{c}{Inner Wall} &  \multicolumn{1}{c}{}& \multicolumn{4}{c}{Outer Wall}\\
\cline{4-7} 
\cline{9-12}
     & Type &   & a$_{max}$  & T$_{wall}^i$ &z$_{wall}^i$ & R$_{wall}^i$   & & a$_{max}$  & T$_{wall}^o$ &z$_{wall}^o$         & R$_{wall}^o$ \\
     &  &  & ({\micron})   &(K)                  &(AU)               & (AU)                  & & ({\micron})   &(K)                    &(AU)                        & (AU)                 \\
~~~~(1)     &(2)  &  & (3)   &(4)                 &(5)           & (6)                 & & (7)  &(8)                   &(9)                      & (10)                \\
\hline
FM~177		& TD             & & -- & -- & -- & -- & & 0.25	& 90	& 3.1 & 49  \\
FM~281		& TD             & & -- & -- & -- & -- & & 0.25	& 90	& 3.6  & 31   \\
LRLL~67		& TD             & & -- & -- & -- & -- & & 5.0	& 130& 1.5  & 10   \\
LRLL~72		& TD            & & -- & -- & -- & -- & & 1.0	& 190& 0.6  & 5  \\
LRLL~133		& TD		  & & -- & -- & -- & -- & & 0.25  & 180 & 0.6 & 4   \\	 
FM~515		& PTD             & & 10 & 1700$^{a}$ & 0.0071 & 0.12 & & 0.25 & 150& 5  & 45  \\
FM~618		& PTD             & & 1.0 & 1400 & 0.0055 & 0.22 & & 5.0	& 180& 0.7  & 11   \\
LRLL~21		& PTD             & & 2 & 1800$^{b}$ & 0.0017 & 0.13 & & 2.0	& 220 & 0.9  & 9   \\
LRLL~31		& PTD               & & 1.0 & 1400 & 0.01 & 0.32 & & 5.0	& 180& 1.5  & 14   \\
LRLL~37		& PTD               & & 0.25 & 1400 & 0.0098 & 0.17 & & 0.25	& 240& 0.6  & 5   \\
FM~581		& FD             & & 0.25 & 1400 & 0.016 & 0.3 & & -- & -- & -- & -- \\
LRLL~2		& FD             & & 0.25 & 1400 & 0.013 & 1.68 & & -- & -- & -- & --   \\
LRLL~6		& FD             & & 1.0 & 1400 & 0.005 & 0.54 & & -- & -- & -- & --   \\
LRLL~55		& FD             & & 0.25 & 1400 & 0.02 & 0.14 & & -- & -- & -- & --   \\
LRLL~68		& FD            & & 10.0 & 1400 & 0.0023 & 0.07 & & -- & -- & -- &~~~~~~ --  
\tablecomments{Col~(1): Name of target. Col~(2): Assigned classification for our targets.  We label objects as transitional disks (TD), pre-transitional disks (PTD), and full disks (FD). Col~(3): Maximum grain size of dust used for the inner wall of the disk.  The superscript $i$ denotes ``inner wall." Col~(4): Temperature of the inner wall. Col~(5): Height of the inner wall. Col~(6): Radius of the inner wall. Col~(7): Maximum grain size of dust used for the outer wall of the disk.  The superscript $o$ denotes ``outer wall." Col~(8): Temperature of the outer wall. Col~(9): Height of the outer wall. Col~(10): Radius of the outer wall.}
\tablenotetext{a}{For FM~515, the inner wall model required a temperature of 1700~K  in order to fit
the slope of the NIR emission.  In some cases, temperatures $>$1400~K for the
inner wall have also been needed to fit the SED previously \citep[e.g. T35, UX Tau
A; ][E11]{espaillat10}.}
\tablenotetext{b}{We adopt a temperature of 1800~K for the inner wall based upon NIR SpeX spectral fitting \citet{flaherty12}.}
\enddata
\end{deluxetable}

\begin{deluxetable}{lcccccccc}
\tabletypesize{\scriptsize}
\tablewidth{0pt}
\tablecaption{Model Runs for LRLL~68\label{tab:ic68}}
\startdata
\hline
\hline
\multicolumn{1}{c}{Model} & \multicolumn{1}{c}{$i$} & \multicolumn{1}{c}{$R_d$}  & \multicolumn{1}{c}{} & 
$\epsilon$ &$\alpha$ & M$_{disk}$ \\
     &  & (AU)  &  &                   &                &(M$_{\sun}$)\\
\hline
1		& 60           & 100 &  & 0.0001 & 0.0006 &  0.005  \\
2		& 60            & 100&    & 0.001 & 0.006 &  0.0005   \\
3		& 60             & 100 &    & 0.01 & 0.06 &  0.00004   \\
4		& 60             & 300 &    & 0.001 & 0.006 &  0.001   \\
5		& 60            & 20 &   & 0.001 & 0.006 &  0.0001   \\
6		& 20		  & 100 &    & 0.001 & 0.006 &  0.001 \\	 
7		& 40           & 100 &   & 0.001 & 0.006 &  0.001  \\
8		& 80             & 100&   & 0.001 & 0.006 &  0.001
\enddata
\end{deluxetable}

\begin{deluxetable}{lccccc}
\tabletypesize{\scriptsize}
\tablewidth{0pt}
\tablecaption{Mass Fraction (in $\%$) of Silicates in Outer Wall in Transitional and Pre-transitional Objects\label{tab:silwall}}
\startdata
\hline
\hline
Target & Amorphous  & Amorphous & Crystalline  &  Crystalline   & Crystalline  \\
            & Olivine           &  Pyroxene     &        Forsterite           &     Enstatite              & Silica  \\
 \hline
FM~177		& 80  & -- & 10 & -- & 10 \\
FM~281		& 90  & -- & -- & -- & 10 \\
FM~515		& 100  & -- & -- & -- & -- \\
FM~618		& 100  & -- & -- & -- & -- \\
LRLL~21		& 80  & -- & 20 & -- & -- \\
LRLL~31		& 80  & -- & -- & 20 & -- \\
LRLL~37		& 70   & -- & 20 & 10 & -- \\
LRLL~67		& 100  & -- & -- & -- & -- \\
LRLL~72		& 60  & 30 & 5 & 5 & -- \\
LRLL~133		& 90  & -- & 10 & -- & -- 
\enddata
\end{deluxetable}

\begin{deluxetable}{lccccc}
\tabletypesize{\scriptsize}
\tablewidth{0pt}
\tablecaption{Mass Fraction (in $\%$) of Silicates in Full Disks \label{tab:silwall2}}
\startdata
\hline
\hline
Target & Amorphous  & Amorphous & Crystalline  &  Crystalline   & Crystalline  \\
            & Olivine           &  Pyroxene     &        Forsterite           &     Enstatite              & Silica  \\
 \hline
FM~581		& 60  & -- & 20 & 10 & 10 \\
LRLL~2		& 60  & -- & 20 & 10 & 10 \\
LRLL~6		& 50  & -- & 20 & 10 & 20 \\
LRLL~55		& 80  & -- & 10 & 5 & 5 \\
LRLL~68		& 85  & -- & 5 & 5 & 5
\enddata
\end{deluxetable}

\begin{deluxetable}{lccccc}
\tabletypesize{\scriptsize}
\tablewidth{0pt}
\tablecaption{Mass Fraction (in $\%$) of Silicates in Optically Thin Dust Region in Transitional and Pre-transitional Objects \label{tab:silthin}}
\startdata
\hline
\hline
Target & Amorphous  & Amorphous & Crystalline  &  Crystalline   & Crystalline  \\
            & Olivine           &  Pyroxene     &        Forsterite           &     Enstatite              & Silica  \\
 \hline
FM~177		& 20  & -- & 40 & 40 & -- \\
FM~281		& 80  & -- & 20 & -- & -- \\
FM~515		& 100  & -- & -- & -- & -- \\
FM~618		& 100  & -- & -- & -- & -- \\
LRLL~21		& 50   & -- & 20 & -- & 30 \\
LRLL~31		& --  & -- & -- & -- & -- \\
LRLL~37		& 40   & -- & 20 & 20 & 20 \\
LRLL~67		& 95   & -- & 5 & -- & -- \\
LRLL~72		& --  & -- & -- & -- & -- \\
LRLL~133		& 90  & -- & 10 & -- & -- 
\enddata
\tablecomments{For LRLL~31, we did not need an optically thin region to account for the 10~{\micron} silicate emission.  This feature comes from the inner wall which has 60$\%$ amorphous olivine and 40$\%$ crystalline forsterite.  
 } 
\end{deluxetable}

\begin{deluxetable}{lccccc}
\tabletypesize{\scriptsize}
\tablewidth{0pt}
\tablecaption{Properties of Optically Thin Dust Region in Transitional and Pre-transitional Objects \label{tab:thinprop}}
\startdata
\hline
\hline
Target & Organics (\%)  & Troilite (\%) & Silicates (\%)  &  M$_{dust}$ (10$^{-12}${\msun})   & a$_{max}$ ({\micron})  \\
 \hline
FM~177		& 42  & 11 & 47 & 2 & 5 \\
FM~281		& 19  & 15 & 66 & 0.9 & 0.25 \\
FM~515		& 19  & 15 & 66 & 0.8 & 0.25 \\
FM~618		& 12  & 9 & 79 & 2 & 3 \\
LRLL~21		& 19   & 15 & 66 & 0.6 & 2 \\
LRLL~31		& --  & -- & -- & -- & -- \\
LRLL~37		& 26   & 14 & 60 & 7 & 5 \\
LRLL~67		& 42   & 11 & 47 & 6 & 5 \\
LRLL~72		& --  & -- & -- & -- & -- \\
LRLL~133		& 42  & 11 & 47 & 6 & 0.25 
\enddata
\tablecomments{We use use a minimum grain size of 0.005 {\micron} in the optically thin dust region.  
 } 
\end{deluxetable}

\clearpage

\clearpage

\clearpage

\clearpage

\clearpage

\clearpage

\clearpage

\clearpage

\clearpage

\clearpage

\clearpage

\clearpage

\clearpage

\clearpage

\end{document}